\documentclass{article}    

\usepackage{latexsym}     
\usepackage{ams}     
\usepackage{amsthm}  
\usepackage{amsmath}     

\newtheorem{theorem}{Theorem}
\newtheorem{lemma}{Lemma}
\newtheorem*{claim}{Claim}

\newtheorem{remark}{Remark}

\theoremstyle{definition}
\newtheorem{definition}{Definition}
\newtheorem{notation}{Notation}

\newcommand{\cC}{\ensuremath{\mathcal{C}}}   
\newcommand{\cE}{\ensuremath{\mathcal{E}}}   
\newcommand{\cF}{\ensuremath{\mathcal{F}}}   
\newcommand{\cI}{\ensuremath{\mathcal{I}}}   
\newcommand{\cL}{\ensuremath{\mathcal{L}}}   
\newcommand{\cG}{\ensuremath{\mathcal{G}}} 
\newcommand{\cGC}{\ensuremath{\mathcal{GC}}} 
\newcommand{\cQ}{\ensuremath{\mathcal{Q}}} 
\newcommand{\cFQ}{\ensuremath{\mathcal{FQ}}} 
\newcommand{\cS}{\ensuremath{\mathcal{S}}} 
\newcommand{\cFS}{\ensuremath{\mathcal{FS}}} 

\newcommand{\fA}{\ensuremath{\mathfrak{A}}}  
\newcommand{\fB}{\ensuremath{\mathfrak{B}}}  
\newcommand{\fD}{\ensuremath{\mathfrak{D}}}  

\title{Data-Complexity of the Two-Variable Fragment with Counting
Quantifiers} \author{Ian Pratt-Hartmann}
\date{}
\begin{document}
\maketitle

\begin{abstract}
\noindent
The data-complexity of both satisfiability and finite satisfiability
for the two-variable fragment with counting is NP-complete; the
data-complexity of both query-answering and finite query-answering for
the two-variable guarded fragment with counting is co-NP-complete.

\vspace{0.2cm}

\noindent
{\bf Keywords: } data-complexity, query answering, two-variable
fragment with counting.
\end{abstract}
\section{Introduction}
\label{sec:intro}
Let $\varphi$ be a sentence (i.e.~a formula with no free variables) in
some logical fragment, $\psi(\bar{y})$ a formula with free variables
$\bar{y}$, $\Delta$ a set of ground, function-free literals, and
$\bar{a}$ a tuple of individual constants with the same arity as
$\bar{y}$. We are to think of $\Delta$ as being a body of {\em data},
$\varphi$ a {\em background theory}, and $\psi(\bar{a})$ a {\em query}
which we wish to answer.  That answer should be positive just in case
$\Delta \cup \{\varphi\}$ entails $\psi(\bar{a})$.  What is the
computational complexity of our task?

A fair reply depends on what, precisely, we take the inputs to our
problem to be. For, in practice, the background theory $\varphi$ is
{\em static}, and the query $\psi(\bar{y})$ {\em small}: only the
database $\Delta$, which is devoid of logical complexity, is large and
indefinitely extensible. Accordingly, we define the {\em
query-answering problem with respect to} $\varphi$ and $\psi(\bar{y})$
as follows: given a set $\Delta$ of ground, function-free literals and
a tuple $\bar{a}$ of individual constants with the same arity as
$\bar{y}$, determine whether $\Delta \cup \{\varphi\}$ entails
$\psi(\bar{a})$.  Similarly, we define the {\em finite query answering
problem with respect to} $\varphi$ and $\psi(\bar{y})$ as follows:
given $\Delta$ and $\bar{a}$, determine whether $\Delta \cup
\{\varphi\}$ entails $\psi(\bar{a})$ under the additional assumption
that the domain of quantification is finite.  The computational
complexity of (finite) query-answering problems is typically lower
than that of the corresponding entailment problem in which all the
components are treated, on a par, as input.  From a theoretical point
of view, it is natural to consider the special case where
$\psi(\bar{y})$ is the falsum. Taking complements, we define the {\em
satisfiability problem with respect to} $\varphi$ as follows: given a
set $\Delta$ of ground, function-free literals, determine whether
$\Delta \cup \{\varphi\}$ is satisfiable. Likewise, we define the {\em
finite satisfiability problem with respect to} $\varphi$ is as
follows: given $\Delta$, determine whether $\Delta \cup \{\varphi\}$
is finitely satisfiable.

The complexity of these problems depends, of course, on the logical
fragments to which $\varphi$ and $\psi(\bar{y})$ are assumed to
belong.  It is common practice to take $\psi(\bar{y})$ to be a {\em
positive conjunctive query}---that is, a formula of the form $\exists
\bar{x} \pi(\bar{x},\bar{y})$, where $\pi(\bar{x},\bar{y})$ is a
conjunction of atoms featuring no function-symbols. This restriction
is motivated by the prevalence of database query-languages, such as,
for example, SQL, in which the simplest and most natural queries have
precisely this form.  By contrast, the choice of logical fragment for
$\varphi$ is much less constrained: in principle, it makes sense to
consider almost any set of formulas for this purpose.  Once we have
identified a logic $\cL$ from which to choose $\varphi$, we can obtain
bounds on the complexity of the (finite) satisfiability problem and
the (finite) query answering problem with respect to any sentence
$\varphi$ in $\cL$ and any positive conjunctive query $\psi(\bar{y})$.
These complexity bounds are collectively referred to as {\em data
complexity bounds} for $\cL$.

In this paper, we analyse the data complexity of two expressive
fragments of first-order logic for which the complexity of
satisfiability and finite satisfiability has recently been determined:
the two-variable fragment with counting quantifiers, denoted $\cC^2$,
and the two-variable guarded fragment with counting quantifiers,
denoted $\cGC^2$. We show that the satisfiability and finite
satisfiabilty problems with respect to any $\cC^2$-formula are in NP,
and that the query-answering and finite query-answering problems with
respect to any $\cGC^2$-formula and any positive conjunctive query are
in co-NP. We show that these bounds are the best possible, and that
the query-answering and finite query-answering problems with respect
to a $\cC^2$-formula and a positive conjunctive query are in general
undecidable. The data complexity of various logical fragments with
counting quantifiers has been investigated in the literature (see, for
example, Hustadt {\em et al.}~\cite{logic:hms2007}, Glimm {\em et
al.}~\cite{logic:g+al08}, Ortiz{\em et al.}~\cite{logic:oce08}, and
Artale {\em et al.}~\cite{logic:ackz07}). However, this is the first
time that such results have been established for the large (and
mathematically natural) fragments $\cC^2$ and $\cGC^2$. In addition,
the proofs in this paper are based ultimately on the technique of
reduction to Presburger arithmetic, which is novel in this context.

\section{Preliminaries}
\label{sec:sor}
We employ the standard apparatus of first-order logic (assumed to
contain the equality predicate $\approx$) augmented with the {\em
counting quantifiers}, $\exists_{\leq C}$, $\exists_{\geq C}$ and
$\exists_{= C}$ (for $C \geq 0$), which we interpret in the obvious
way. The {\em predicate calculus with counting}, denoted $\cC$, is the
the set of first-order formulas with counting quantifiers, over a
purely relational signature.  The {\em two-variable fragment with
counting}, denoted $\cC^2$, is the fragment of $\cC$ involving only
the variables $x$ and $y$, and only unary or binary predicates.  If
$r$ is any binary predicate (including $\approx$), we call an atomic
formula having either of the forms $r(x,y)$ or $r(y,x)$ a {\em
guard}. Note that guards, by definition, contain two distinct
variables.  The {\em two variable guarded fragment with counting},
denoted $\cGC^2$, is the smallest set of formulas satisfying the
following conditions:
\begin{enumerate}
\item $\cGC^2$ contains all atomic formulas, and is closed under Boolean
combinations;
\item if $\varphi$ is a formula of $\cGC^2$ with at most one free
variable, and $u$ is a variable (i.e.~either $x$ or $y$), then the
formulas $\forall u \varphi$ and $\exists u \varphi$ are in $\cGC^2$;
\item if $\varphi$ is a formula of $\cGC^2$, $\gamma$ a guard, $u$ a
variable, and $Q$ any of the quantifiers $\exists$, $\exists_{\leq
C}$, $\exists_{\geq C}$, $\exists_{= C}$ (for $C >0$), then the
formulas $\forall u (\gamma \rightarrow \varphi)$, $Q u (\gamma \wedge
\varphi)$ and $Q u \gamma$ are in $\cGC^2$.
\end{enumerate}
For example, 
\begin{equation}
\exists_{\leq 1} x ({\rm professor}(x) \wedge 
   \exists_{\geq 4} y ({\rm supervises}(x,y) \wedge {\rm grad\_student}(y)))
\label{eq:C2ex}
\end{equation}
is a $\cC^2$-sentence, with the informal reading: {\sf At most one
professor supervises more than three graduate students}. Likewise,
\begin{equation*}
\neg \exists x ({\rm professor}(x) \wedge 
   \exists_{\geq 41} y ({\rm supervises}(x,y) \wedge {\rm grad\_student}(y)))
\end{equation*}
is a $\cGC^2$-sentence, with the informal reading: {\sf No professor
supervises more than forty graduate students}.
However,~\eqref{eq:C2ex} is not in the fragment $\cGC^2$, because the
quantifier $\exists_{\leq 1}$ does not occur in a guarded pattern.  It
will be convenient in the sequel to consider the following smaller
fragments. We take $\cL^{2-}$ to be the fragment of $\cC^2$ in which
no counting quantifiers and no instances of $\approx$ occur; likewise, we take
$\cG^{2-}$ to be the fragment of $\cGC^2$ in which no counting
quantifiers and no instances of $\approx$ occur. Evidently, $\cG^{2-} \subseteq
\cL^{2-}$.

Both $\cC^2$ and $\cGC^2$ lack the finite model property.  The
satisfiability and finite satisfiability problems for $\cC^2$ are both
NEXPTIME-complete (Pratt-Hartmann \cite{logic:ph05}; see also
Pacholski~{\em et al.}~\cite{logic:pst99}); the satisfiability and
finite satisfiability problems for $\cGC^2$ are both EXPTIME-complete
(Kazakov~\cite{logic:kazakov04}, Pratt-Hartmann~\cite{logic:ph07a}).
In the context of $\cC^2$ and $\cGC^2$, predicates of arities other
than 1 or 2 lead to no interesting increase in expressive power.
Adding individual constants to $\cC^2$ likewise leads to no
interesting increase in expressive power, and no increase in
complexity, since occurrences of any constant $c$ can be simulated
with a unary predicate $p_c$ in the presence of the $\cC^2$-formula
$\exists_{=1} x p_c(x)$. On the other hand, adding even a single
individual constant to $\cGC^2$ results in a fragment with
NEXPTIME-complete satisfiability and finite-satisfiability problems.
Thus, it is most convenient to assume these fragments to be
constant-free; and that is what we shall do in the sequel.

A {\em positive conjunctive query} (or, simply: {\em query}) is a
formula $\psi(\bar{y})$ of the form $\exists \bar{x} \left(
\alpha_1(\bar{x}, \bar{y}) \wedge \cdots \wedge \alpha_n(\bar{x},
\bar{y}) \right)$, where $n \geq 1$ and, for all $i$ ($1 \leq i \leq
n$), $\alpha_1(\bar{x}, \bar{y})$ is an atomic formula whose predicate
is not $\approx$, and whose arguments are all variables occurring in
$\bar{x}, \bar{y}$. Since we shall be interested in answering queries
in the presence of $\cC^2$- or $\cGC^2$-formulas, there is little to
be gained from allowing $\psi(\bar{y})$ to contain predicates of arity
greater than 2; in the sequel, therefore, we assume that all
predicates in positive conjunctive queries are unary or binary.  An
{\em instance} of $\psi(\bar{y})$ is simply the corresponding formula
$\psi(\bar{a})$, where $\bar{a}$ is a tuple of constants.  We allow
the tuples $\bar{x}$ and $\bar{y}$ to be empty. Allowing individual
constants to appear in positive conjunctive queries does not
essentially change the problem; in the sequel, therefore, we assume
positive conjunctive queries to be constant-free.

\begin{definition}
If $\varphi$ is a sentence (in any logic), define $\cS_\varphi$ to be the
following problem:
\begin{quote}
Given a finite set of ground, function-free literals $\Delta$,
is $\Delta \cup \{ \varphi \}$ satisfiable?
\end{quote}
Likewise, define 
$\cFS_\varphi$ to be the
following problem:
\begin{quote}
Given a finite set of ground, function-free literals $\Delta$, is
$\Delta \cup \{ \varphi \}$ finitely satisfiable?
\end{quote}
\label{def:sat}
We call $\cS_\varphi$ the {\em satisfiability problem with respect to}
$\varphi$, and $\cFS_\varphi$ the {\em finite satisfiability problem
with respect to} $\varphi$.
\end{definition}
\begin{definition}
If $\varphi$ is a sentence and $\psi(\bar{y})$ a formula (in any
logic) having no free variables apart from $\bar{y}$, define
$\cQ_{\varphi, \psi(\bar{y})}$ to be the following problem:
\begin{quote}
Given a finite set of ground, function-free literals $\Delta$ and a
tuple of constants $\bar{a}$ of the same arity as $\bar{y}$, does
$\Delta \cup \{ \varphi \}$ entail $\psi(\bar{a})$?
\end{quote}
Likewise, define 
$\cFQ_{\varphi,
\psi(\bar{y})}$
to be the following
problem:
\begin{quote}
Given a finite set of ground, function-free literals $\Delta$ and a
tuple of constants $\bar{a}$ of the same arity as $\bar{y}$, is
$\psi(\bar{a})$ true in every finite model of $\Delta \cup \{ \varphi
\}$?
\end{quote}
\label{def:queryAnswering}
We call $\cQ_{\varphi,\psi(\bar{y})}$ the {\em query answering problem
with respect to} $\varphi$ and $\psi(\bar{y})$, and
$\cFQ_{\varphi,\psi(\bar{y})}$ the {\em finite query answering problem
with respect to} $\varphi$ and $\psi(\bar{y})$.
\end{definition}
\noindent
Answering queries is at least as hard as deciding {\em
un}satisfiability: if $p$ is any predicate not occurring in $\Delta$
or $\varphi$, then then $\Delta \cup \{\varphi\} \models \exists x
p(x)$ if and only if $\Delta \cup \{\varphi\}$ is
unsatisfiable. Similarly for the finite case.

We establish the following complexity results.  For any
$\cC^2$-sentence $\varphi$, both $\cS_\varphi$ and $\cFS_\varphi$ are in NP.
These bounds are tight in the sense that there exists a
$\cC^2$-sentence---in fact, a $\cG^{2-}$-sentence---$\varphi$ such that the
problems $\cS_\varphi$ and $\cFS_\varphi$ coincide, and are are NP-hard.
The query-answering problem for $\cC^2$ is of little interest from a
complexity-theoretic point of view: there exist a $\cC^2$-sentence
$\varphi$ and a positive conjunctive query $\psi(\bar{y})$ such that
$\cQ_{\varphi, \psi(\bar{y})}$ is undecidable; similarly for $\cFQ_{\varphi,
\psi(\bar{y})}$.  However, by restricting attention to, $\cGC^2$, we
restore upper complexity bounds comparable to those for $\cS_\varphi$ and
$\cFS_\varphi$: for any $\cGC^2$-sentence $\varphi$ and any positive
conjunctive query $\psi(\bar{y})$, both $\cQ_{\varphi, \psi(\bar{y})}$
and $\cFQ_{\varphi, \psi(\bar{y})}$ are in co-NP.  Again, the fact that
there exists a $\cG^{2-}$-sentence $\varphi$ for which $\cS_\varphi (=
\cFS_\varphi)$ is NP-hard means that these bounds are tight.  The
above results may be informally expressed by saying: ``The
data-complexity of (finite) satisfiability for $\cC^2$ is NP-complete;
the data-complexity of (finite) query-answering for $\cGC^2$ is
co-NP-complete.''  These data-complexity bounds contrast with the
complexity bounds for satisfiability and finite satisfiability in the
fragments $\cC^2$ and $\cGC^2$ mentioned above.

In the sequel, if $\varphi$ is a formula, $\lVert \varphi \rVert$ denotes the
size of $\varphi$, measured in the obvious way; similarly, if $\varphi$ is a
set of formulas, $\lVert \varphi \rVert$ denotes the total size of
$\varphi$. If $X$ is any set, $|X|$ denotes the cardinality of $X$.

\section{The fragment $\cC^2$}
\label{sec:C2}
In this section, we review some facts about the fragment $\cC^2$,
closely following the analysis in Pratt-Hartmann~\cite{logic:ph05}. We
have simplified the original terminology where, for the purposes of
the present paper, certain complications regarding the sizes of
data-structures can be disregarded; and we have lightly reformulated
some of the lemmas accordingly.

Let $\Sigma$ be a signature of unary and binary predicates.  A 1-{\em
type} over $\Sigma$ is a maximal consistent set of equality-free
literals involving only the variable $x$. A 2-{\em type} over $\Sigma$
is a maximal consistent set of equality-free literals involving only
the variables $x$ and $y$.  If $\fA$ is any structure interpreting
$\Sigma$, and $a \in A$, then there exists a unique 1-type $\pi(x)$
over $\Sigma$ such that $\fA \models \pi[a]$; we denote $\pi$ by ${\rm
tp}^\fA[a]$. If, in addition, $b \in A$ is distinct from $a$, then
there exists a unique 2-type $\tau(x,y)$ over $\Sigma$ such that $\fA
\models \tau[a,b]$; we denote $\tau$ by ${\rm tp}^\fA[a,b]$.  We do
not define ${\rm tp}^\fA[a,b]$ if $a = b$.  If $\pi$ is a 1-type, we
say that $\pi$ is {\em realized} in $\fA$ if there exists $a \in A$
with ${\rm tp}^\fA[a] = \pi$.  If $\tau$ is a 2-type, we say that
$\tau$ is {\em realized} in $\fA$ if there exist distinct $ a, b \in
A$ with ${\rm tp}^\fA[a,b] = \tau$.
\begin{notation} 
  Let $\tau$ be a 2-type over a purely relational signature $\Sigma$.
  The result of transposing the variables $x$ and $y$ in $\tau$ is
  also a 2-type, denoted $\tau^{-1}$; the set of literals in $\tau$
  not featuring the variable $y$ is a 1-type, denoted ${\rm
  tp}_1(\tau)$; likewise, the set of literals in $\tau$ not featuring
  the variable $x$ is also a 1-type, denoted ${\rm tp}_2(\tau)$.
\label{notation:2:types}
\end{notation}
\begin{remark}
If $\tau$ is any 2-type over a purely relational signature $\Sigma$,
then ${\rm tp}_2(\tau) = {\rm tp}_1(\tau^{-1})$. If $\fA$ is a
structure interpreting $\Sigma$, and $a$, $b$ are distinct elements of
$A$ such that ${\rm tp}^\fA[a,b] = \tau$, then ${\rm tp}^\fA[b,a] =
\tau^{-1}$, ${\rm tp}^\fA[a] = {\rm tp}_1(\tau)$ and ${\rm tp}^\fA[b]
= {\rm tp}_2(\tau)$.
\label{remark:types}
\end{remark}
\begin{lemma}
Let $\varphi$ be a $\cC^2$-formula.  There exist $($i$)$ a
$\cC^2$-formula $\alpha$ containing no quantifiers and no occurrences
of $\approx$, $($ii$)$ a list of positive integers $C_1, \ldots, C_m$
and $($iii$)$ a list of binary predicates $f_1, \ldots, f_m$, with the
following property. If $\varphi^*$ is the $\cC^2$-formula
\begin{equation}
\forall x \forall y (\alpha \vee x \approx y) \wedge 
   \bigwedge_{1 \leq h \leq m} 
      \forall x \exists_{=C_h} y (f_h(x,y) \wedge x \not \approx y),
\label{eq:normalFormC2}
\end{equation}
and $C = \max_h C_h$, then $($i$)$ $\varphi^* \models \varphi$, and $($ii$)$
any model of $\varphi$ over a domain having at least $C+1$ elements may be
expanded to a model of $\varphi^*$.
\label{lma:normalFormC2}
\end{lemma}
\begin{proof}
Routine adaptation of standard techniques. See, e.g.~B{\"o}rger {\em
et al.}~\cite{logic:BGG97}, p.~378.
\end{proof}
If $\Delta$ is a set of ground, function-free literals, and $\varphi$ and
$\varphi^*$ are as in Lemma~\ref{lma:normalFormC2}, then $\Delta \cup
\{\varphi\}$ evidently has a (finite) model if and only if either $\Delta
\cup \{\varphi\}$ has a model of size $C$ or less, or $\Delta \cup
\{\varphi^*\}$ has a (finite) model. 

Lemma~\ref{lma:normalFormC2} assures us that formulas of the
form~\eqref{eq:normalFormC2} are as general as we need. So, for the
remainder of this section, let us fix a formula $\varphi^*$ given
by~\eqref{eq:normalFormC2}.  The predicates $f_1, \ldots, f_m$ will
play a special role in the ensuing analysis. We refer to them as the
{\em counting predicates}. However, we stress that no special
assumptions are made about them: in particular, they can occur in
arbitrary configurations in the sub-formula $\alpha$.

Fix the constant $Z = (mC+1)^2$.  Let $\Sigma^*$ be the signature of
$\varphi^*$ together with $2 \lceil \log Z \rceil +1$ new unary
predicates (i.e.~not occurring in $\varphi^*$). Henceforth, $\Sigma^*$
will be implicit: thus, unless otherwise indicated, {\em structure }
means ``structure interpreting $\Sigma^*$''; 1-{\em type} means
``1-type over $\Sigma^*$''; 2-{\em type} means ``2-type over
$\Sigma^*$''; and so on.
\begin{definition}
  Let $\tau$ be a 2-type.  We say that $\tau$ is a {\em message-type}
  if $f_h(x,y) \in \tau$ for some $h$ ($1 \leq h \leq m$).  If $\tau$
  is a message-type such that $\tau^{-1}$ is also a message-type, we
  say that $\tau$ is {\em invertible}. On the other hand, if $\tau$ is
  a 2-type such that neither $\tau$ nor $\tau^{-1}$ is a message-type,
  $\tau$ is a {\em silent} 2-type.  If $\tau$ is a 2-type such that
  neither $q(x,y)$ nor $q(y,x)$ is in $\tau$ for any binary predicate
  $q$, $\tau$ is {\em vacuous}.
\label{def:messagetype}
\end{definition}
The terminology is meant to suggest the following imagery.  Let $\fA$
be a structure. If ${\rm tp}^\fA[a,b]$ is a message-type $\mu$, then
we may imagine that $a$ sends a message (of type $\mu$) to $b$. If
$\mu$ is invertible, then $b$ replies by sending a message (of type
$\mu^{-1}$) back to $a$. If ${\rm tp}^\fA[a,b]$ is silent, then
neither element sends a message to the other.  Note that every vacuous
2-type is by definition silent; but the converse is not generally
true.

For convenience, we decide upon some enumeration 
\begin{equation*}
\pi_1, \ldots, \pi_L
\end{equation*}
of the set of all 1-types, and  some enumeration 
\begin{equation*}
\mu_1, \ldots, \mu_{M^*}, \mu_{M^*+1}, \ldots, \mu_M
\end{equation*}
of the set of all message-types, such that $\mu_j$ is invertible if $1
\leq j \leq M^*$, and non-invertible if $M^*+1 \leq j \leq M$. (That
is: the invertible message-types are listed first.)  In addition, let
$\Xi$ denote the set of silent 2-types.  The above notation, which
will be used throughout this section, is summarized in
Table~\ref{table:4:symbols}.
\begin{table}
\begin{center}
\begin{tabular}{l|l}
Symbol & Definition\\
\hline
\\
$Z$ & $(mC +1)^2$\\
$\Sigma^*$ & signature of $\varphi$ together with 
$2 \lceil \log Z \rceil +1$ new
unary predicates\\
$\pi_1, \ldots, \pi_L$ & an enumeration of the 1-types over $\Sigma^*$\\
$\mu_1, \ldots, \mu_{M^*}$ & an enumeration of the invertible message-types over
$\Sigma^*$\\
$\mu_{M^*+1}, \ldots, \mu_M$ & an enumeration of 
the non-invertible message-types over
$\Sigma^*$\\
$\Xi$ & set of silent 2-types over $\Sigma^*$
\end{tabular}
\end{center}
\caption{Quick reference guide to symbols defined with respect to 
Formula~\eqref{eq:normalFormC2}.}
\label{table:4:symbols}
\end{table}

We now introduce two notions necessary to state the key lemmas of this
section regarding the satisfiability of $\cC^2$-formulas.
\begin{definition}
  A structure $\fA$ is {\em chromatic} if distinct elements connected
  by a chain of 1 or 2 invertible message-types have distinct
  1-types. That is, $\fA$ is chromatic just in case, for all $a, a',
  a'' \in A$:
\begin{enumerate}
\item if $a \neq a'$ and ${\rm tp}^\fA[a,a']$ is an invertible
  message-type, then ${\rm tp}^\fA[a] \neq {\rm tp}^\fA[a']$; and
\item if $a, a', a''$ are pairwise distinct and both ${\rm
  tp}^\fA[a,a']$ and ${\rm tp}^\fA[a',a'']$ are invertible
  message-types, then ${\rm tp}^\fA[a] \neq {\rm tp}^\fA[a'']$.
\end{enumerate}
\label{def:4:chromatic}
\end{definition}
\noindent
\begin{remark}
A structure is chromatic if and only if \textup{(}i\textup{)} no
object sends an invertible message to any object having the same
1-type as itself; and \textup{(}ii\textup{)} no object sends
invertible messages to any two objects having the same 1-type as each
other.
\label{remark:chromatic}
\end{remark}
\begin{definition}
  A structure $\fA$ is {\em differentiated} if, for every 1-type
  $\pi$, the number $u$ of elements in $A$ having 1-type $\pi$
  satisfies either $u \leq 1$ or $u > Z$.
\label{def:4:differentiated}
\end{definition}

By the L\"{o}wenheim-Skolem Theorem, we may confine attention in the
sequel to finite or countably infinite structures.  The following
(routine) lemma ensures that we may further confine attention to
chromatic, differentiated structures of these cardinalities.
\begin{lemma}
  Suppose $\fA \models \varphi^*$. Then, by re-interpreting $2\lceil \log
  Z \rceil$ of the $2 \lceil \log Z \rceil +1$ unary predicates of $\Sigma^*$ not
  occurring in $\varphi^*$ if necessary, we can obtain a chromatic,
  differentiated structure $\fA'$ over the same domain, such that
  $\fA' \models \varphi^*$.
\label{lma:4:chromaticDifferentiated}
\end{lemma}
\begin{proof}
Pratt-Hartmann~\cite{logic:ph05}, Lemmas~2 and~3.
\end{proof}
 
In the sequel, we shall need to record the cardinalities of various
finite or countably infinite sets.  To this end, we let $\N^* = \N
\cup \{ \aleph_0 \}$, and we extend the ordering $>$ and the
arithmetic operations $+$ and $\cdot$ from $\N$ to $\N^*$ in the
obvious way. Specifically, we define $\aleph_0 > n$ for all $n \in
\N$; we define $\aleph_0 + \aleph_0 = \aleph_0 \cdot \aleph_0 =
\aleph_0$ and $0 \cdot \aleph_0 = \aleph_0 \cdot 0 = 0$; we define $n
+ \aleph_0 = \aleph_0 +n = \aleph_0$ for all $n \in \N$; and we define
$n \cdot \aleph_0 = \aleph_0 \cdot n = \aleph_0$ for all $n \in \N$
such that $n >0$.  Under this extension, $>$ remains a total order,
and $+$, $\cdot$ remain associative and commutative. 

Our next task is to develop the means to talk about `local
configurations' in structures. 
\begin{definition}
  A {\em star-type} is a pair $\sigma = \langle
  \pi, \bar{v} \rangle$, where $\pi$ is a 1-type, and
  $\bar{v} = (v_1, \ldots, v_M)$ is an $M$-tuple over $\N^*$
  satisfying the condition that, for all $j$ ($1 \leq j \leq M$),
\begin{equation*}
v_j >0 \mbox{ implies } {\rm tp}_1(\mu_j) = \pi.  
\end{equation*}
In this context, we denote $\pi$ by ${\rm tp}(\sigma)$ and $v_j$ by
$\sigma[j]$.  If
$\fA$ is a finite or countably infinite structure, and $a \in A$, we
denote by ${\rm st}^\fA[a]$ the star-type $\langle \pi, (v_1, \ldots,
v_M) \rangle$, where $\pi = {\rm tp}^\fA[a]$ and
\begin{equation*}
v_j = 
  | \{ b \in A \setminus \{ a \} : {\rm tp}^\fA[a,b] = \mu_j \} |
\end{equation*}
for all $j$ ($1 \leq j \leq M$). We call ${\rm st}^\fA[a]$ the {\em
star-type of} $a$ {\em in} $\fA$; and we say that a star-type $\sigma$
is {\em realized} in $\fA$ if $\sigma = {\rm st}^\fA[a]$ for some $a
\in A$. 
\label{def:4:star-type}
\end{definition}
We may think of ${\rm st}^\fA[a]$ as a description of the `local
environment' of $a$ in $\fA$: it records, in addition to the 1-type of
$a$ in $\fA$, the number of other elements to which $a$ sends a
message of type $\mu_j$, for each message-type $\mu_j$.  Properties of
star-types realized in models capture `local' information about those
models.
\begin{definition}
Let $\sigma = \langle \pi, (v_1, \ldots, v_M) \rangle$ be a
star-type. We say that $\sigma$ is {\em $D$-bounded}, for $D$ a
positive integer, if $\sigma[j] \leq D$ for all $j$ ($1 \leq j \leq
M$).  We say that $\sigma$ is {\em chromatic} if, for every 1-type
$\pi'$, the sum
\begin{equation*}
c = \sum \{ v_j \mid 1 \leq j \leq M^* \mbox{ and }
                           {\rm tp}_2(\mu_j) = \pi'\}
\end{equation*}
satisfies $c \leq 1$, and satisfies $c = 0$ if $\pi' = \pi$.  We say
that a finite or countably infinite structure $\fA$ is $D$-{\em
bounded} if every star-type realized in $\fA$ is $D$-bounded.
\label{def:startypeCharacteristics}
\end{definition}
Obviously, if $\fA \models \varphi^*$, then $\fA$ is
$C$-bounded. Importantly, information about the populations of
star-types realized in models can tell us all that we need to know
about those models, from the point of view of the fragment $\cC^2$.
\begin{definition}
Let $\fA$ be a finite or countably infinite structure, and let
$\bar{\sigma} = \sigma_1, \ldots, \sigma_N$ be a list of
star-types. 
For all $k$ ($1 \leq
k \leq N$), let $w_k \in \N^*$ be given by
\begin{equation*}
w_k = |\{ a \in A \mid {\rm st}^\fA[a] = \sigma_k \}|.
\end{equation*}
The $\bar{\sigma}$-{\em histogram} of $\fA$, denoted
$H_{\bar{\sigma}}(\fA)$, is the $N$-tuple $(w_1, \ldots, w_N)$.
\label{def:historam}
\end{definition}

We may thus think of $H_{\bar{\sigma}}(\fA)$ as a `statistical profile' of
$\fA$. For the next definitions, recall (Table~\ref{table:4:symbols})
that $\pi_1, \ldots, \pi_L$, is an enumeration of the 1-types, and
that $\Xi$ is the set of silent 2-types.

\begin{definition}
If $\fA$ is a structure and $\pi$, $\pi'$ are 1-types (not necessarily
distinct), we say that $\pi$ and $\pi'$ {\em form a quiet pair} in
$\fA$ if there exist distinct elements $a$ and $a'$ of $A$, such that
${\rm tp}[a] = \pi$, ${\rm tp}[a'] = \pi'$ and ${\rm tp}[a,a']$ is
silent.
\label{def:quiet}
\end{definition}
\begin{definition}  
  Let $\cI$ be the set of unordered pairs of $($not necessarily
  distinct$)$ integers between $1$ and $L$: that is, $\cI = \{
  \{i,i'\} \mid 1 \leq i \leq i' \leq L\}$.  A {\em frame} is a triple
  $\cF = (\bar{\sigma}, I, \theta)$, satisfying:
  \begin{enumerate}
  \item $\bar{\sigma} = (\sigma_1, \ldots, \sigma_N)$ is an  
  $N$-tuple of pairwise distinct star-types for some $N>0$;
  \item $I \subseteq \cI$; and 
  \item $\theta: I \rightarrow \Xi$ is a function such that, for all
  $\{i, i'\} \in I$ with $i \leq i'$, ${\rm tp}_1(\theta(\{i,i'\}))
  = \pi_i$ and ${\rm tp}_2(\theta(\{i,i'\})) = \pi_{i'}$.
  \end{enumerate}
The frame $\cF$ is $D$-{\em bounded} if every star-type in
$\bar{\sigma}$ is $D$-bounded.  Likewise, $\cF$ is {\em chromatic} if
every star-type in $\bar{\sigma}$ is chromatic.
\label{def:4:frame}
\end{definition}
Think of a frame $\cF = (\bar{\sigma}, I, \theta)$ as a (putative)
schematic description of a structure, where $\bar{\sigma}$ tells us
which star-types are realized, $I$ tells us which pairs of 1-types are
quiet, and $\theta$ selects, for each quiet pair of 1-types, a silent
2-type joining them. More precisely:
\begin{definition}
  Let $\fA$ be a structure and $\cF = (\bar{\sigma}, I, \theta)$ a
  frame.  We say that $\cF$ {\em describes} $\fA$ if the following
  conditions hold:
\begin{enumerate}
\item $\bar{\sigma}$ is a list of all and only those star-types
realized in $\fA$;
\item if $\pi_i$ and $\pi_{i'}$ form a quiet pair in $\fA$, then
$\{ i, i' \} \in I$;
\item if $\pi_i$ and $\pi_{i'}$ form a quiet pair in $\fA$, then there
exist distinct $a, a' \in A$ such that ${\rm tp}^\fA[a,a'] = \theta(\{
i, i' \})$.
\end{enumerate}
\label{def:4:framefor}
\end{definition}
\noindent

Frames contain the essential information required to determine whether
certain structures they describe are models of $\varphi^*$.  The next
definition employs the notation established in
Table~\ref{table:4:symbols} and Definition~\ref{def:4:star-type}.
\begin{definition}
We write $\cF \models \varphi^*$ if the following conditions are
satisfied:
\begin{enumerate}
\item for all $k$ ($1 \leq k \leq N$) and all $j$ ($1 \leq j \leq M$),
  if $\sigma_k[j] > 0$ then \linebreak
  $\models \bigwedge \mu_j \rightarrow
  \alpha(x,y) \wedge \alpha(y,x)$;
\label{item:4:fladescription1}
\item for all $\{i,i'\} \in I$, $\models \bigwedge \theta(\{i,i'\})
  \rightarrow \alpha(x,y) \wedge \alpha(y,x)$;
\label{item:4:fladescription2}
\item for all $k$ ($1 \leq k \leq N$) and all $h$ ($1 \leq h \leq m$),
the sum of all the $\sigma_k[j]$ ($1 \leq j \leq M$) such that
$f_h(x,y) \in \mu_j$ equals $C_h$.
\label{item:4:fladescription3}
\end{enumerate}
\label{def:frameModels}
\end{definition}
The next lemma helps to motivate this definition.
\begin{lemma}
  If $\fA \models \varphi^*$, then there exists a frame $\cF$ describing
  $\fA$, such that $\cF \models \varphi^*$.
\label{lma:4:flaModel}
\end{lemma} 
The proof is almost immediate: Conditions 1 and 2 in
Definition~\ref{def:frameModels} are secured by the fact that $\fA
\models \forall x \forall y (\alpha \vee x \approx y)$, while
Condition 3 is secured by the fact that $\fA \models \bigwedge_{1 \leq
h \leq m} \forall x \exists_{=C_h} y (f_h(x,y) \wedge x \not \approx
y)$.  The following Lemma also follows almost immediately from the
above definitions.
\begin{lemma}
  Let $\fA$ be a structure, $\cF$ a frame describing $\fA$, and $D$ a 
positive integer. Then:
\begin{enumerate}
\item $\cF$ is $D$-bounded if and only if $\fA$ is $D$-bounded;
\item $\cF$ is chromatic if and only if $\fA$ is chromatic;
\end{enumerate}
\label{lma:4:fladescription}
\end{lemma} 

However, while every structure is described by some frame, not every
frame describes a structure; and it is important for us to define a
class of frames which do. To this end, we associate with a frame $\cF$
a collection of numerical parameters, as follows.
\begin{notation}
  Let $\cF = (\bar{\sigma}, I, \theta)$ be a frame, where
  $\bar{\sigma} = (\sigma_1, \ldots, \sigma_N)$, for some $N>0$, and
  recall the notation established in Table~\ref{table:4:symbols} and
  Definition~\ref{def:4:star-type}. If $\cF$ is clear from context,
  for integers $i,k$ in the ranges $1 \leq i \leq L$, $1 \leq k \leq
  N$ write:
\begin{align*}
o_{ik} & =
\begin{cases}
1 \mbox{ if ${\rm tp}(\sigma_k) = \pi_i$}\\
0 \mbox{ otherwise;}
\end{cases}\\
p_{ik} &= 
\begin{cases}
1 \mbox{ if, for all $j$ $($$1 \leq j \leq M$$)$, 
           ${\rm tp}_2(\mu_{j}) = \pi_i$ implies $\sigma_k[j]= 0$}\\
0 \mbox{ otherwise;}
\end{cases}\\
r_{ik} &= \sum_{j \in J} \sigma_k[j], 
 \mbox{ where }  J = \{j \mid M^*+ 1 \leq j \leq M \mbox{ and }  
                                   {\rm tp}_2(\mu_j) = \pi_i \};\\
s_{ik} &= \sum_{j \in J} \sigma_k[j], 
 \mbox{ where }  J = \{j \mid 1 \leq j \leq M \mbox{ and } 
                                   {\rm tp}_2(\mu_j) = \pi_i \}.
\end{align*}
In addition, for integers $i,j$ in the ranges $1 \leq i \leq L$,
$1 \leq j \leq M^*$, write:
\begin{align*}
q_{jk} &= \sigma_k[j].
\end{align*}
\label{notation:4:description}
\end{notation}

With this notation in hand we can characterize a class of frames whose
members are guaranteed to describe structures.
\begin{definition}
  Let $\cF = (\bar{\sigma}, I, \theta)$ be a frame,
  where $\bar{\sigma} = (\sigma_1, \ldots, \sigma_N)$. Let $\bar{w} =
  (w_1, \ldots, w_N)$ be an $N$-tuple over $\N^*$.  Using
  Notation~\ref{notation:4:description}, for all $i$ ($1 \leq i \leq
  L$), all $i'$ ($1 \leq i' \leq L$) and all $j$ ($1 \leq j \leq
  M^*$), let:
\begin{equation*}
u_i = \sum_{1 \leq k \leq N} o_{ik} w_{k} \hspace{1cm}
v_j = \sum_{1 \leq k \leq N} q_{jk} w_{k} \hspace{1cm}
x_{ii'} = \sum_{1 \leq k \leq N} o_{ik}p_{i'k} w_{k}.
\end{equation*}
We say that an $N$-tuple $\bar{w}$
over $\N^*$ is a {\em solution} of $\cF$ if the
following conditions are satisfied for all $i$ ($1 \leq i \leq L$),
all $i'$ ($1 \leq i' \leq L$), all $j$ ($1 \leq j \leq M^*$) and all
$k$ ($1 \leq k \leq N$):
\begin{description}
\item[(C1) ] $v_j = v_{j'}$, where $j'$ is such that $\mu_j^{-1}
  = \mu_{j'}$;
\item[(C2) ] $s_{ik} \leq u_i$;
\item[(C3) ] $u_i \leq 1$ or $u_i >Z$;
\item[(C4) ] if $o_{ik} =1$, then either $u_i > 1$ or $r_{i'k} \leq
  x_{i'i}$;
\item[(C5) ] if $\{ i, i' \} \not \in I$, then either $u_i \leq 1$ or
  $u_{i'} \leq 1$;
\item[(C6) ] if $\{ i, i' \} \not \in I$ and $o_{ik} =1$, then 
  $r_{i'k} \geq x_{i'i}$.
\end{description}
\label{def:4:finitelyZsolvable}
\end{definition}

The conditions~{\bf C1}--{\bf C6} in
Definition~\ref{def:4:finitelyZsolvable} may be written as a
quantifier-free formula in the language of Presburger arithmetic---in
other words, as a Boolean combination of linear inequalities with
integer coefficients and variables $w_1, \ldots, w_N$.  By treating a
negated inequality as a reversed inequality in the obvious way, we may
assume that the Boolean combination in question is {\em
positive}---i.e.~ involves only conjunction and disjunction. Denote
this positive Boolean combination of inequalities by $\cE$. By
definition, $\cF$ has a solution if and only if $\cE$ is satisfied
over $\N^*$; and $\cF$ has a finite solution (i.e.~a solution in which
all values are finite) if and only if $\cE$ is satisfied over $\N$.

We are at last in a position to state the key lemmas of this
section.
\begin{lemma}
  If $\fA$ is a differentiated structure and $\cF = \langle
  \bar{\sigma}, I, \theta \rangle$ is a frame
  describing $\fA$, then $H_{\bar{\sigma}}(\fA)$ is a
  solution of $\cF$.
\label{lma:4:maineasy}
\end{lemma}
\begin{proof}
Pratt-Hartmann~\cite{logic:ph05}, Lemma~13, Lemma~16.
\end{proof}
\begin{lemma}
  If $\cF$ is a chromatic frame such that $\cF \models \varphi^*$, and
  $\bar{w}$ is a solution of $\cF$, then there exists a structure
  $\fA$ such that: \textup{(}i\textup{)} $\fA \models \varphi^*$;
  \textup{(}ii\textup{)} $\cF$ describes $\fA$; and
  \textup{(}iii\textup{)} $\bar{w} = H_{\bar{\sigma}}(\fA)$.
\label{lma:4:mainhard}
\end{lemma}
\begin{proof}
Pratt-Hartmann~\cite{logic:ph05}, Lemma~14, Lemma~17.
\end{proof}

Lemmas~\ref{lma:4:maineasy} and~\ref{lma:4:mainhard} in effect state
that, to determine the satisfiability of $\varphi^*$, it suffices to
guess a $C$-bounded, differentiated, chromatic frame $\cF$, and to
test that $\cF$ has a solution and that $\cF \models
\varphi^*$. Furthermore, by testing instead whether $\cF$ has a {\em
finite} solution, we can determine the {\em finite} satisfiability of
$\varphi^*$. The proof of Lemma~\ref{lma:4:maineasy} is relatively
straightforward; that of Lemma~\ref{lma:4:mainhard} is more
challenging, because it involves constructing a model $\fA$ of
$\varphi^*$, given only the frame $\cF$ and its solution. It can in
fact be shown that we may without loss of generality confine attention
to frames whose size (measured in the obvious way) is bounded by a
singly exponential function of the size of $\varphi^*$
(Pratt-Hartmann~\cite{logic:ph05}, Lemma~10). From this it follows
that the problems of determining the satisfiability/finite
satisfiability of a given $\cC^2$-formula are in NEXPTIME. In the
present context of investigating the {\em data}-complexity of $\cC^2$,
however, this matter may be safely ignored.

\section{Data-complexity of satisfiability and finite satisfiability for $\cC^2$} 
\label{sec:dataC2}
In this section, we give bounds on the data-complexity of
satisfiability and finite satisfiability in $\cC^2$.

We consider the upper bounds first. For any $\cC^2$-formula $\varphi$, we
describe a pair of non-deterministic polynomial-time procedures to
determine the satisfiability and finite satisfiability of $\Delta \cup
\{ \varphi \}$, where $\Delta$ is a given set of ground, non-functional
literals. The strategy is as follows. Relying on
Lemmas~\ref{lma:4:maineasy} and~\ref{lma:4:mainhard}, we guess a frame
$\cF$ such that $\cF \models \varphi$, and assemble the inequalities
required for $\cF$ to have a solution.  By augmenting these
inequalities with extra conditions (based on $\Delta$), we can check
for the existence of a (finite) model of $\varphi$ whose histogram (with
respect to some sequence of star-types) is such that a model of
$\Delta$ can be spliced into it, thus yielding a model of $\Delta \cup
\{ \varphi \}$.

If $\Delta$ is a set of ground, function-free literals, we denote by
${\rm const}(\Delta)$ the set of individual constants occurring in
$\Delta$.
\begin{theorem}
For any $\cC^2$-sentence $\varphi$, both $\cS_\varphi$ and $\cFS_\varphi$ are
in {\rm NP}.
\label{theo:dataSatC2}
\end{theorem}
\begin{proof}
Let $\varphi$ be a $\cC^2$-formula, and $\Delta$ a set of ground,
function-free literals, over a signature $\Sigma_\Delta$.  Let
$\varphi^*$ and $C$ be as in Lemma~\ref{lma:normalFormC2}. Determining
whether $\Delta \cup \{\varphi\}$ has a model of size $C$ or less is
straightforward.  For we may list, in constant time, all models of
$\varphi$ of size $C$ or less (interpreting the signature of
$\varphi$).  Fixing any such model $\fA$, we may then guess an
expansion $\fA^+$ of $\fA$ interpreting $\Sigma_\Delta$, and check
that $\fA^+ \models \Delta$.  This (non-deterministic) process can be
executed in time bounded by a linear function of $\lVert \Delta
\rVert$.  Hence, it suffices to determine whether $\Delta \cup
\{\varphi^*\}$ has a model.

From now on, we fix the formula $\varphi^*$ having the
form~\eqref{eq:normalFormC2}, and employ the notation of
Table~\ref{table:4:symbols}, together with the associated notions of
{\em 1-type}, {\em message-type} and {\em star-type} over the
signature $\Sigma^*$. Since $\Sigma^*$ contains $2\lceil \log Z \rceil
+1$ unary predicates not occurring in $\varphi$, pick one of these extra
predicates, $o$. We call a 1-type $\pi$ {\em observable} if $o(x) \in
\pi$, we call a message-type $\rho$ {\em observable} if ${\rm
tp}_1(\rho)$ and ${\rm tp}_2(\rho)$ are observable, and we call a
star-type $\sigma$ {\em observable} if ${\rm tp}(\sigma)$ is
observable.  Informally (and somewhat approximately), we read $o(x)$
as ``$x$ is an element which interprets a constant in $\Delta$''.

We now define two non-deterministic procedures operating on $\varphi^*$
and $\Delta$. We show that both procedures run in time bounded by a
polynomial function of $\lVert \Delta \rVert$, that the first of these
procedures has a successful run if and only if $\Delta \cup
\{\varphi^*\}$ is satisfiable, and that the second has a successful run
if and only if $\Delta \cup \{\varphi^*\}$ is finitely satisfiable. This
proves the theorem.  Procedure~I is as follows.
\begin{enumerate}
\item Guess a structure $\fD^+$ interpreting the signature $\Sigma^*
  \cup \Sigma_\Delta$ over a domain $D$ with $|D| \leq {\rm
  const}(\Delta)$; and let $\fD$ be the reduct of $\fD^+$ to the
  signature $\Sigma^*$. If $\fD^+ \not \models \Delta$ or $\fD \not
  \models \forall x \forall y (\alpha \vee x \approx y)$, then fail.
\label{enum:step1}
\item 
Guess a list $\sigma_1, \ldots, \sigma_{N'}$ of
observable, $C$-bounded, chromatic  star-types, and guess a further list
$\sigma_{N'+1}, \ldots, \sigma_N$ of non-observable, $C$-bounded,
chromatic star-types. Write
\begin{equation*}
\bar{\sigma} = \sigma_1, \ldots, \sigma_{N'}, \sigma_{N'+1}, \ldots, \sigma_N,
\end{equation*}
and guess a frame $\cF = \langle \bar{\sigma}, I, \theta \rangle$ with these 
star-types. If $\cF \not \models \varphi^*$, then fail.
\label{enum:step2} \item Guess a function $\delta: D \rightarrow \{\sigma_1, \ldots,
\sigma_{N'}\}$ mapping every element of $D$ to one of the observable
star-types of $\cF$. Writing $\langle \pi^d, (v_1^d, \ldots, v_M^d)
\rangle$ for $\delta(d)$, if, for any $d \in D$, either of the
conditions
\begin{enumerate}
\item $\pi^d = {\rm tp}^\fD[d]$
\label{enum:subStep3a}
\item for all $j$ ($1 \leq j \leq M$) such that $\rho_j$ is an
observable message-type, 
\begin{equation*}
v_j^d = |\{d' \in D \mid d' \neq d \mbox{ and } {\rm tp}^\fD[d,d'] = \mu_j \}|
\end{equation*}
\label{enum:subStep3b}
\end{enumerate}

\vspace{-0.75cm}

does not hold, then fail. Otherwise, record the numbers $n_1, \ldots,
n_{N'}$, where, for all $k$ ($1 \leq k \leq N'$), $n_k =
|\delta^{-1}(\sigma_k)|$, and then forget $\delta$.
\label{enum:step3}
\item Let $\cE$ be the (positive) Boolean combination of inequalities
required for $\cF$ to have a solution, as explained in
Section~\ref{sec:C2}. Guess the truth-values of all the inequalities
involved in $\cE$. If the guess makes $\cE$ false (considered as a
Boolean combination), fail; otherwise, let $\cE'$ be the set of these
inequalities guessed to be true.
\label{enum:step4}
\item Recalling the numbers $n_k$ from Step~\ref{enum:step3} let
\begin{equation*}
\cE'_\delta = \cE' \cup \{w_k = n_k \mid 1 \leq k \leq N' \}. 
\end{equation*}
If there is no solution of $\cE'_\delta$, then fail.
\label{enum:step5}
\item Succeed.
\label{enum:step6}
\end{enumerate}
Procedure II is exactly the same as Procedure I, except in
Step~\ref{enum:step5}.  Instead of failing if there is no solution of
$\cE'_\delta$, we instead fail if there is no finite solution of
$\cE'_\delta$.

\vspace{0.25cm}

\noindent
We consider the running time of Procedure I, writing $\lVert \Delta
\rVert = n$. Step~\ref{enum:step1} can be executed in time $O(n^3)$.
Step~\ref{enum:step2} can be executed in constant time.  In executing
Step~\ref{enum:step3}, we note that, once $\delta(d)$ has been guessed
and checked, the space required to do so can be recovered; only the
tallies $n_1, \ldots, n_{N'}$ need be kept, and this never requires
more than $N'\log n$ space.  Moreover, in checking $\delta(d)$, the
only difficulty is to compute the quantities $|\{d' \in D \mid d' \neq
d \mbox{ and } {\rm tp}^\fD[d,d'] = \mu_j \}|$ for observable message
types $\mu_j$; but this never requires more than $\log n$ space. Hence
Step~\ref{enum:step3}, can be executed in space $O(\log(n))$, and
hence in time bounded by a polynomial function of $n$.
Step~\ref{enum:step4} can be executed in constant time.
Step~\ref{enum:step5} involves determining the existence of a solution
to the inequalities in $\cE'_\delta$. Since the size of $\cE'$ is
bounded by a constant, the size of $\cE'_\delta$ is in fact $O(\log
n)$; moreover, $\cE'_\delta$ involves a fixed number of
variables. After guessing which of these variables take infinite
values, this problem can be solved using Lenstra's algorithm
(Lenstra~\cite{logic:lenstra83}) in time bounded by some fixed
polynomial function of $\log n$, and hence certainly in time $O(n)$.
Thus, Procedure I can be executed in polynomial time.  Procedure II
can also be executed in polynomial time, by an almost identical
argument.

\vspace{0.25cm}

\noindent
We show that Procedure I has a successful run if and only if $\Delta
\cup \{ \varphi^* \}$ is satisfiable, and that Procedure II has a
successful run if and only if $\Delta \cup \{ \varphi^* \}$ is finitely
satisfiable.  Suppose $\fA^+$ is a finite or countably infinite model
of $\Delta \cup \{ \varphi^* \}$, interpreting the signature $\Sigma^*
\cup \Sigma_\Delta$ over a domain $A$; let $\fA$ be the reduct
of $\fA^+$ to $\Sigma^*$; and let $D \subseteq A$ be the set of all
and only those elements interpreting the constants ${\rm
const}(\Delta)$ in $\fA^+$.  By assumption, $\Sigma^*$ contains
$2\lceil \log Z \rceil +1$ unary predicates not occurring in $\varphi^*$,
one of which is the predicate $o$. By re-interpreting these new
predicates if necessary, we may assume that $o^\fA = D$, and
furthermore (by Lemma~\ref{lma:4:chromaticDifferentiated}) that $\fA$
is differentiated and chromatic.  Let $\fD^+$ be the restriction of
$\fA^+$ to $D$, and $\fD$ the restriction of $\fA$ to $D$ (so that
$\fD$ is a reduct of $\fD^+$).  With these choices,
Step~\ref{enum:step1} succeeds.  By Lemma~\ref{lma:4:flaModel}, let
$\cF$ be a frame describing $\fA$ such that $\cF \models \varphi^*$. By
Lemma~\ref{lma:4:fladescription}, Parts 1 and 2, $\cF$ is $C$-bounded
and chromatic. Without loss of generality, we may assume the
star-types in $\cF$ to be $\bar{\sigma} = \sigma_1, \ldots,
\sigma_{N'}, \sigma_{N'+1}, \ldots, \sigma_N$, where $\sigma_1,
\ldots, \sigma_{N'}$ are the star-types realized in $\fA$ by elements
of $D$, and $\sigma_{N'+1}, \ldots, \sigma_N$ are the star-types
realized in $\fA$ be elements of $A \setminus D$. With these choices,
Step~\ref{enum:step2} succeeds.  Define $\delta: D \rightarrow
\{\sigma_1, \ldots, \sigma_{N'}\}$ by setting $\delta(d) = {\rm
st}^\fA[d]$.  With these choices, Step~\ref{enum:step3} succeeds. Let
$\bar{w} = H_{\bar{\sigma}}(\fA)$, so that, by
Lemma~\ref{lma:4:maineasy}, $\bar{w}$ is a solution of $\cE$.  Let
$\cE'$ be the set of inequalities mentioned in $\cE$ which are
satisfied by $\bar{w}$.  With these choices, Step~\ref{enum:step4}
succeeds.  The above choice of $\bar{w}$ ensures that $\bar{w}$
satisfies $\cE'$; to show that Step~\ref{enum:step5}---and hence the
whole procedure---succeeds, it suffices to show that, for all $k$ ($1
\leq k \leq N'$) $w_k = n_k$. Now, since $o^\fA = D$, $a \in A$ has an
observable star-type $\sigma_k$ if and only if $a \in D$. But for $d
\in D$, we have $\delta(d) = {\rm st}^\fA[d]$, whence $n'_k =
|\delta^{-1}(\sigma_k)|$ is the number of elements $d \in D$ such that
${\rm st}^\fA[d] = \sigma_k$, and hence the number of elements $a \in
A$ such that ${\rm st}^\fA[a] = \sigma_k$.  That is: $w_k = n_k$ as
required. The corresponding argument for Procedure II is almost
identical, noting that, if $\fA^+$ is finite, then $\bar{w} =
H_{\bar{\sigma}}(\fA)$ will consist entirely of finite values.

\vspace{0.25cm}

\noindent
Suppose, conversely, that Procedure I has a successful run. Let
$\fD^+$, $\fD$, $\delta$, $\cF$, and $\cE'$ be as guessed in this run,
and let $\bar{w} = w_1, \ldots, w_N$ be a solution of $\cE'_\delta$,
guaranteed by the fact that Step~\ref{enum:step5} succeeds.  Since
Step~\ref{enum:step1} succeeds, we have $\fD^+ \models \Delta$, and
$\fD \models \forall x \forall y (\alpha \vee x \approx y)$. By
assumption, $\cF$ is chromatic; moreover, since Step~\ref{enum:step2}
succeeds, $\cF \models \varphi^*$. Since Step~\ref{enum:step4}
succeeds, $\bar{w}$ is a solution of the Boolean combination of
inequalities $\cE$, and hence a solution of the frame $\cF$. By
Lemma~\ref{lma:4:mainhard}, then, let $\fA$ be a model of $\varphi^*$
described by $\cF$ in which the star-types $\sigma_1, \ldots,
\sigma_N$ are realized $w_1, \ldots, w_N$ times, respectively.

We proceed to define a structure $\fA'$ such that $\fA' \models \Delta
\cup \{ \varphi^* \}$.  Let $D' = o^\fA$, and, for all $k$ ($1 \leq k
\leq N'$), let $D'_k = \{a \in A \mid {\rm st}^\fA[a] = \sigma_k\}$.
Evidently, the sets $D'_1, \ldots, D'_{N'}$ partition $D'$.  On the
other hand, consider the domain $D$ of the structure $\fD$, and, for
all $k$ ($1 \leq k \leq N'$), let $D_k = \delta^{-1}(\sigma_k)$. These
sets are pairwise disjoint, and from the fact that $\bar{w}$ is a
solution of $\cE'_\delta$, we have $|D_k| = |D'_k|$, for all $k$ ($1
\leq k \leq N'$). By replacing $\fA$ with a suitable isomorphic copy
if necessary, we can assume that $D_k = D'_k$ for all $k$ ($1 \leq k
\leq N'$). We thus have: (i) $D = D' \subseteq A$; (ii) ${\rm
st}^\fA[d] = \delta(d)$ for all $d \in D$; and (iii) $o^\fA = D$.  Now
define the structure $\fA'$ interpreting $\Sigma^*$ over the domain
$A$ by setting:
\begin{equation*}
{\rm tp}^{\fA'}[a,b] = 
\begin{cases}
{\rm tp}^{\fD}[a,b] \text{ if $a \in D$ and $b \in D$}\\
{\rm tp}^{\fA}[a,b] \text{ otherwise.}
\end{cases}
\end{equation*}
To ensure that no clashes can occur in these assignments, we must show
that ${\rm tp}^\fA[a] = {\rm tp}^\fD[a]$ for all $a \in D$.  But this
follows from the success of Step~\ref{enum:step3} (specifically, from
Condition~\ref{enum:subStep3a}) and the already-established fact that
${\rm st}^\fA[a] = \delta(a)$.  By construction, then, $\fD \subseteq
\fA'$. Indeed, taking $\fA^+$ to be the expansion of $\fA'$ obtained
by interpreting the symbols of $\Sigma_\Delta \setminus \Sigma^*$ in
the same way as $\fD^+$, we immediately have $\fA^+ \models \Delta$.
To show that $\Delta \cup \{\varphi^*\}$ is satisfiable, therefore, we
require only to show that $\fA' \models \varphi^*$. Note first of all
that the only 2-types realized in $\fA'$ are 2-types realized either
in $\fA$ or in $\fD$. But $\fA \models \varphi^*$, and $\fD \models
\forall x \forall y (\alpha \vee x \approx y)$, whence $\fA' \models
\forall x \forall y (\alpha \vee x \approx y)$. Therefore, it suffices
to show that, for all $a \in A$, ${\rm st}^{\fA'}[a] = {\rm
st}^\fA[a]$, from which it follows that $\fA' \models \bigwedge_{1
\leq h \leq m} \forall x \exists_{=C_h} y (f_h(x,y) \wedge x \not
\approx y)$.  If $a \not \in D$, then ${\rm st}^{\fA'}[a] = {\rm
st}^\fA[a]$ is immediate from the construction of $\fA'$; so suppose
$a = d \in D$. Let us write
\begin{eqnarray*}
{\rm st}^\fA[d] = \delta(d) & = & \langle \pi, (v^d_1, \ldots, v^d_M) \rangle\\
{\rm st}^{\fA'}[d] & = & \langle \pi, (v'_1, \ldots, v'_M) \rangle.
\end{eqnarray*}
Fix $k$ ($1 \leq j \leq M$), and suppose first that $\rho_j$ is not
observable. Since $D \subseteq o^\fA$, we have, by the construction of
$\fA'$, ${\rm tp}^{\fA'}[d,b] = \mu_j$ if and only if $b \not \in D$
and ${\rm tp}^\fA[d,b] = \mu_j$; it is then immediate that $v'_j =
v^d_j$.  Suppose, on the other hand, that $\rho_j$ is
observable. Since $o^\fA \subseteq D$, we have, by the construction
of $\fA'$ ${\rm tp}^{\fA'}[d,b] = \mu_j$ if and only if $b \in D$ and
${\rm tp}^\fD[d,b] = \mu_j$; but then the success of
Step~\ref{enum:step3} (specifically, Condition~\ref{enum:subStep3b})
then guarantees that $v'_j = v^d_j$. Hence, for all $a \in A$, ${\rm
st}^{\fA'}[a] = {\rm st}^\fA[a]$, as required.  The corresponding
argument for Procedure II is almost identical: we need only observe
that, by requiring the numbers $w_{N'+1}, \ldots, w_N$ to be in $\N$,
the constructed model $\fA^+$ will be finite.
\end{proof}

The matching lower bound to Theorem~\ref{theo:dataSatC2} is almost
trivial. In fact, much smaller fragments than $\cC^2$ suffice for this
purpose: recall that $\cG^{2-}$ is the fragment of $\cGC^2$ in which
no counting quantifiers and no instances of $\approx$ occur.
\begin{theorem}
There exists a $\cG^{2-}$-sentence $\varphi$ for which the problems
$\cS_\varphi$ and $\cFS_\varphi$ coincide, and are {\rm NP}-hard.
\label{theo:dataSatC2hard}
\end{theorem}
\begin{proof}
By reduction of 3SAT.
Let $c$ and $t$ be unary predicates and $l_1$, $l_2$, $l_3$, $o$ and
$s$ binary predicates. (Read $c(x)$ as ``$x$ is a clause'', $l_i(x,y)$
as ``$y$ is the $i$th literal of $x$'', $t(x)$ as ``$x$ is a true
literal'', $o(x,y)$ as ``$x$ and $y$ are mutually opposite literals'', and
$s(x,y)$ as ``$x$ and $y$ are the same literal''.)  Let $\varphi$ be
\begin{multline*}
 \forall x (c(x) \rightarrow 
   \bigvee_{1 \leq j \leq 3} 
       \exists y(l_j(x,y) \wedge t(y)))
         \wedge \\
   \forall x \forall y (o(x,y) \rightarrow (t(x) \leftrightarrow \neg t(y)))
       \wedge 
  \forall x \forall y (s(x,y) \rightarrow (t(x) \leftrightarrow t(y)))
    \wedge \\
\bigwedge_{1 \leq j \leq 3} 
    \forall x(\exists y(l_j(x,y) \wedge t(y)) \rightarrow
       \forall y(l_j(x,y) \rightarrow t(y))). 
\end{multline*}
We reduce 3SAT to the problems $\cS_\varphi$ and $\cFS_\varphi$, which we
simultaneously show to be identical.  Suppose a finite set $\Gamma =
\{C_1, \ldots, C_n\}$ of 3-literal clauses is given, where $C_i =
L_{i,1} \vee L_{i,2} \vee L_{i,3}$.  Let $a_i$ ($1 \leq i \leq n$) and
$b_{i,j}$ ($1 \leq i \leq n$; $1 \leq j \leq 3$) be pairwise 
distinct individual
constants, and let $\Delta_\Gamma$ be the following set of ground,
function-free literals:
\begin{multline*}
\{c(a_i) \mid 1 \leq i \leq n\} \cup
\{ l_j(a_i,b_{i,j}) \mid 1 \leq i \leq n \mbox{ and } 1 \leq j \leq 3 \} \cup\\
 \{ o(b_{i,j},b_{i',j'}) \mid \mbox{$L_{i,j} \mbox{ and }
L_{i',j'}$ are opposite literals}\}
          \cup \\
 \{ s(b_{i,j},b_{i',j'}) \mid \mbox{$L_{i,j} \mbox{ and }
L_{i',j'}$ are the same literal}\}.
\end{multline*}
It is routine to check that: ({\em i}) if $\{\varphi\} \cup \Delta_\Gamma$
is satisfiable, then $\Gamma$ is satisfiable; ({\em ii}) if $\Gamma$
is satisfiable, then $\{\varphi\} \cup \Delta_\Gamma$ is finitely
satisfiable. 
\end{proof}

Since, as we remarked above, the (finite) query-answering problem is
at least as hard as the (finite) {\em un}satisfiability problem,
Theorem~\ref{theo:dataSatC2hard} also provides a lower bound for the
complexity of (finite) query answering in $\cGC^2$ (matching
Theorem~\ref{theo:queryAnswering} below). Specifically, let $\varphi
\in \cGC^2$ be the sentence constructed in the proof of
Theorem~\ref{theo:dataSatC2hard}, and $p$ a unary predicate; then the
problems $\cQ_{\varphi,\exists x p(x)}$ and $\cFQ_{\varphi, \exists x
p(x)}$ coincide, and are co-NP-complete. We remark that lower
complexity bounds of co-NP for query-answering problems are not always
be obtained in this way (i.e.~by reduction to the corresponding
unsatisfiability problem), especially in inexpressive fragments.  A
good example is provided by the fragments considered in Calvanese {\em
et al.}~\cite{logic:calvanese+al06} (Theorem~8), who use instead a
closely related result on `instance checking' in description logics
(Schaerf~\cite{logic:schaerf93}, Theorem 3.2). For similar results
concerning an expressive logic, see Hustadt {\em et
al.}~\cite{logic:hms2007}, Theorems~20 and~26.

We conclude this section by showing that there is no hope of extending
Theorem~\ref{theo:dataSatC2} to a result concerning query answering:
query-answering and finite query answering problems with respect to
$\cC^2$-formulas are in general undecidable. (Again, much smaller
fragments than $\cC^2$ suffice for this purpose.)  We employ the
standard apparatus of tiling systems.  In this context, recall that a
{\em tiling system} is a triple $T = \langle C, H, V \rangle$, where
$C$ is a non-empty, finite set of {\em tiles} and $H$, $V$ are binary
relations on $C$. For $N \in \N$, let $\N_N$ denote the set $\{0, 1,
\ldots, N-1 \}$. An {\em infinite tiling} for $T$ is a function $f:
\N^2 \rightarrow C$ such that, for all $i, j \in \N$, $\langle f(i,j),
f(i+1,j) \rangle \in H$ and $\langle f(i,j), f(i,j+1) \rangle \in
V$. An $N$-{\em tiling} for $T$ is a function $f: \N_N^2 \rightarrow
C$ such that, for all $i, j \in \N_N$, $\langle f(i,j), f(i+1,j) \rangle
\in H$ and 
$\langle f(i,j), f(i,j+1) \rangle \in V$ (addition modulo $N$). 
The {\em infinite
tiling problem on} $T$ is the following problem: given a sequence
$c_0, \ldots, c_n$ of elements of $C$ (repeats allowed), determine
whether there exists an infinite tiling $f$ for $T$ such that $f(i,0)
= c_i$ for all $i$ ($0 \leq i \leq n$). The {\em finite tiling
problem on} $T$ is the following problem: given a sequence $c_0,
\ldots, c_n$ of elements of $C$ (repeats allowed), determine whether
there exist an $N > n$ and an $N$-tiling $f$ for $T$ such that $f(i,0)
= c_i$ for all $i$ ($0 \leq i \leq n$). It is well-known that there
exist tiling systems for which the infinite tiling problem is
co-r.e.-complete, and that there exist tiling systems for which the
finite tiling problem is r.e.-complete. 
\begin{lemma}
Let $h$ and $v$ be binary predicates, and let $\gamma$ be the formula
\begin{equation*}
\forall x_1 \forall x_2 \forall x_3 \forall x_4 (h(x_1,x_2) \wedge
   v(x_1,x_3) \wedge v(x_2,x_4) \rightarrow h(x_3,x_4)). 
\end{equation*}
There exists a sentence $\varphi$ in $\cG^{2-}$ such that the problem
$\cS_{\varphi \wedge \gamma}$ is co-r.e.-complete.  There exists a
sentence $\varphi$ in $\cG^{2-}$ such that the problem $\cFS_{\varphi \wedge
\gamma}$ is r.e.-complete.
\label{lma:grid}
\end{lemma}
\begin{proof}
Let $T = \langle C, H, V \rangle$ be a tiling system whose infinite tiling
problem is co-r.e.-complete. Treating the tiles $c \in C$ as unary
predicates, let $\varphi_0$ be the formula
\begin{equation*}
\forall x \exists y h(x,y) \wedge \forall x \exists y v(x,y),
\end{equation*}
let $\varphi_T$ be the formula 
\begin{multline*}
\forall x \left( \bigvee_{c \in C} c(x) \right) \wedge 
\bigwedge_{c \neq c'} \forall x (c(x) \rightarrow \neg c'(x)) \wedge \\
 \bigwedge_{\langle c, c' \rangle \not \in H} \forall x \forall y (h(x,y) \rightarrow
      \neg(c(x) \wedge c(y))) \wedge \\
 \bigwedge_{\langle c, c' \rangle \not \in V} \forall x \forall y (v(x,y) \rightarrow
      \neg(c(x) \wedge c(y))),
\end{multline*}
and let $\varphi$ be $\varphi_0 \wedge \varphi_T$. Now, given a sequence
$\bar{c} = c_0, \ldots, c_n$ of elements of $C$ (repeats allowed), let
$a_0, \ldots, a_n$ be individual constants, and let $\Delta_{\bar{c}}$
be the set of ground, function-free literals
\begin{equation*}
\{ c_0(a_0), h(a_0,a_1), c_1(a_1), h(a_1,a_2), \ldots, c_{n-1}(a_{n-1}), 
   h(a_{n-1},a_n), c_n(a_n) \}.
\end{equation*}
We claim that the instance $\bar{c}$ of the infinite tiling problem
for $T$ is positive if and only if $\Delta \cup \{\varphi \wedge
\gamma\}$ is satisfiable.  Thus, the problem $\cS_{\varphi \wedge
\gamma}$ is co-r.e.-complete, proving the first statement of the
lemma.

To prove the claim, if $f$ is an infinite tiling for $T$ with $f(i,0)
= c_i$ for all $i$ ($0 \leq i \leq n$), construct the model $\fA$ as
follows. Let $A = \N^2$; let $a_i^\fA = (i,0)$ for all $i$ ($0 \leq
i \leq n$); let $h^\fA = \{\langle (i,j), (i+1,j) \rangle \mid i, j
\in \N \}$; let $v^\fA = \{\langle (i,j), (i,j+1) \rangle \mid i, j
\in \N \}$; and let $c^\fA = \{(i,j) \mid f(i,j) = c\}$ for all $c \in
C$. It is routine to check that $\fA \models \{\varphi \wedge \gamma
\} \cup \Delta_{\bar{c}}$. Conversely, suppose $\fA \models \{\varphi
\wedge \gamma \} \cup \Delta_{\bar{c}}$. Define a function $g: \N^2
\rightarrow A$ as follows. First, set $g(i,0) = a_i^\fA$ for all $i$
($0 \leq i \leq n$). Now, if $i$ is the largest integer such that
$g(i,0)$ has been defined, select any $b \in A$ such that $\langle
g(i,0), b\rangle \in h^\fA$ (possible, since $\fA \models \varphi_0$),
and set $g(i+1,0) = b$.  This defines $g(i,0)$ for all $i \in
\N$. Fixing any $i$, if $j$ is the largest integer such that $g(i,j)$
has been defined, select any $b \in A$ such that $\langle g(i,j),
b\rangle \in v^\fA$ (possible, since $\fA \models \varphi_0$), and set
$g(i,j+1) = b$.  This defines $g(i,j)$ for all $i,j \in \N$. Since
$\fA \models \Delta_{\bar{c}} \cup \{\gamma\}$, we have, for all $i, j
\in \N$, $\langle (i,j), (i+1,j) \rangle \in h^\fA$ and $\langle
(i,j), (i,j+1) \rangle \in v^\fA$. We now define an infinite tiling
$f: \N^2 \rightarrow C$ as follows. Since $\fA \models \varphi_T$,
we set $f(i,j)$ to be the unique $c \in C$ such that $\fA \models
c[g(i,j)]$. Finally, since $\fA \models \Delta_{\bar{c}}$, we have
$f(i,0) = c_i$ for all $i$ ($1 \leq i \leq n$).

The second statement of the lemma is proved analogously.
\end{proof}
Recall that we denote by $\cL^{2-}$ the fragment of $\cC^2$ in which
no counting quantifiers and no instances of $\approx$ occur.
\begin{theorem}
There exist an $\cL^{2-}$-sentence $\varphi'$ and a positive conjunctive query
$\psi(\bar{y})$ such that $\cQ_{\varphi', \psi(\bar{y})}$ is
undecidable. Similarly for $\cFQ_{\varphi', \psi(\bar{y})}$.
\label{theo:queryAnsweringC2}
\end{theorem}
\begin{proof}
We deal with $\cQ_{\varphi', \psi(\bar{y})}$ only; the proof for
$\cFQ_{\varphi', \psi(\bar{y})}$ is analogous.  Let the binary predicate
$h$ and the formulas $\gamma$ and $\varphi$ be as in (the first
statement of) Lemma~\ref{lma:grid}. Let $p$ be a new unary predicate
and $\bar{h}$ a new binary predicate. Now let $\varphi'$ be the formula
\begin{equation*}
\varphi \wedge \forall x y (\bar{h}(x,y) \leftrightarrow \neg h(x,y)),
\end{equation*}
and $\psi$ the positive conjunctive query
\begin{equation*}
\exists x_1 \exists x_2 \exists x_3 \exists x_4 \exists x (h(x_1,x_2) \wedge
   v(x_1,x_3) \wedge v(x_2,x_4) \wedge \bar{h}(x_3,x_4) \wedge p(x)). 
\end{equation*}
It is obvious that, if $\Delta$ is any set of ground, non-functional
literals (not involving the predicates $p$ or $\bar{h}$), then 
\begin{eqnarray*}
\Delta \cup \{\varphi'\} \models \psi & \mbox{ iff } 
       & \Delta \cup \{\varphi' \wedge \gamma \} \models \exists x p(x) \\
\  & \mbox{ iff } 
       & \Delta \cup \{\varphi' \wedge \gamma \} \mbox{ is unsatisfiable} \\
\  & \mbox{ iff } 
       & \Delta \cup \{\varphi \wedge \gamma \} \mbox{ is unsatisfiable}.
\end{eqnarray*}
It follows from Lemma~\ref{lma:grid} that $\cQ_{\varphi', \psi}$
is undecidable.
\end{proof}

We remark that, at the cost of complicating the above proofs, the
formula $\gamma$ in Lemma~\ref{lma:grid} could in fact have been
replaced by the simpler formula $\forall x_1 \forall x_2 \forall x_3
(r(x_1,x_2) \wedge r(x_2,x_3) \rightarrow r(x_1,x_3))$, asserting the
transitivity of a binary relation. Indeed, it is known that extending
$\cC^2$---or even $\cGC^2$---with the ability to express transitivity
of relations renders the satisfiability problem for this fragment
undecidable.  (Tendera~\cite{logic:tendera05} shows this in the case
of four transitive relations; see also Gr\"{a}del and
Otto~\cite{logic:GO99} for closely related results.)  Notice in this
context that the formula $\varphi'$ constructed in the proof of
Theorem~\ref{theo:queryAnsweringC2} is not in $\cGC^2$, since it
contains the non-guarded conjunct $\forall x y (h(x,y) \leftrightarrow
\neg \bar{h}(x,y))$.  As we shall see in the next section, this is no
accident: query-answering and finite query-answering are decidable
with respect to sentences of $\cGC^2$ and positive conjunctive
queries.  For an investigation of the data-complexity of
satisfiability and query-answering in certain logics featuring both
counting quantifiers and transitive predicates---and indeed of
practical methods for solving these problems---see, for example,
Hustadt {\em et al.}~\cite{logic:hms2007}, Glimm {\em et
al.}~\cite{logic:g+al08}, Ortiz{\em et al.}~\cite{logic:oce08}.

\section{The fragment $\cGC^2$}
\label{sec:GC2}
In this section, we establish some facts about $\cGC^2$ which will
subsequently be used to analyse the complexity of query-answering and
finite query-answering within this fragment. To help motivate this
analysis, we begin with an overview of our approach.

Let $\varphi$ be a sentence of $\cGC^2$, $\Delta$ a set of ground,
function-free literals, and $\varphi(\bar{y})$ a positive conjunctive
query.  For simplicity, let us assume for the moment that the tuple
$\bar{y}$ is empty---that is, $\psi$ is the Boolean query
\begin{equation} 
\exists x_1 \ldots \exists x_n (p_1(y_1, z_1) \wedge \cdots \wedge 
    p_s(y_s,z_s)),
\label{eq:query}
\end{equation}
where the $y_i$ and $z_i$ are chosen from among the set of variables
$V = \{x_1, \ldots, x_n\}$.  Formula~\eqref{eq:query} defines a graph
$(G,E)$ on this set in a natural way: $(x_i,x_j) \in E$ just in case
$i \neq j$ and, for some $k$ ($1 \leq k \leq s$), $\{x_i,x_j\} =
\{y_k,z_k\}$. Again, for simplicity, let us assume for the moment that
the resulting graph, $(V,E)$, is connected.

Now, there are two possibilities: either the graph $(V,E)$ contains a
loop (that is: it is 2-connected) or it does not (that is: it is a
tree).  If the latter, it can be shown (Lemma~\ref{lma:pureTformulas},
below) that $\psi$ is logically equivalent to some $\cGC^2$-formula
$\pi$.  But then the problem $\cQ_{\varphi, \psi}$ is the complement
of the problem $\cS_{\varphi \wedge \neg \pi}$, which is in NP by
Theorem~\ref{theo:dataSatC2}. Suppose, therefore, that $(V,E)$
contains a loop, and consider any model $\fA \models \psi$.  It is
obvious that $\fA$ contains a sequence of elements $a_0, \ldots,
a_{t-1}$ ($t \leq s$) such that for all $i$ ($1 \leq i < t$), there is
a binary predicate $p$ with either $\fA \models p[a_i,a_{i+1}]$ or
$\fA \models p[a_{i+1},a_i]$ (where the addition in the indices is
modulo $t$).  Let us call such a sequence a {\em cycle}.  We therefore
establish the following `big-cycles' lemma for $\cGC^2$-formulas
$\varphi$ (Lemma~\ref{lma:bigLoops}, below): if $\Delta \cup
\{\varphi\}$ is (finitely) satisfiable, then, for arbitrarily large
$\Omega \in \N$, $\Delta \cup \{\varphi\}$ has a (finite) model {\em
in which no cycles with $t \leq \Omega$ exist}. It follows that
$\Delta \cup \{\varphi\}$ is (finitely) satisfiable if and only if
$\Delta \cup \{\varphi, \neg \psi \}$ is (finitely) satisfiable. That
is, the problem $\cQ_{\varphi, \psi}$ is the complement of the problem
$\cS_{\varphi}$, which, again, is in NP by
Theorem~\ref{theo:dataSatC2}; similarly, {\em mutatis mutandis}, for
finite satisfiability.

For satisfiability (as opposed to {\em finite} satisfiability), this
`big-cycles' lemma is relatively straightforward, and close to the
familiar fact that $\cGC^2$ has the `tree-model property' (see
Kazakov~\cite{logic:kazakov04}, Theorem~1). For finite satisfiability,
however, more work is required. We now proceed to lay the foundations
for that work.

\begin{lemma}
Let $\varphi$ be a formula of $\cGC^2$, $\fA$ a structure interpreting
the signature of $\varphi$, and $I$ a nonempty set.  For $i \in I$,
let $\fA_i$ be a copy of $\fA$, with the domains $A_i$ pairwise
disjoint.  If $\varphi$ is satisfied in $\fA$, then it is satisfied in
the structure $\fA'$ with domain $A' = \bigcup_{i \in I} A_i$ and
interpretations $q^{\fA'} = \bigcup_{i \in I} q^{\fA_i}$ for every
predicate $q$.
\label{lma:dup}
\end{lemma}
\begin{proof}
If $\theta: \{x,y\} \rightarrow A$ is any variable assignment, and $i
\in I$, let $\theta_i$ be the variable assignment which maps $x$ and
$y$ to the corresponding elements in $A_i \subseteq A'$. A routine
structural induction on $\varphi$ shows that $\fA \models_\theta \varphi$ if
and only if, for some (= for all) $i \in I$, $\fA' \models_{\theta_i}
\varphi$.
\end{proof}
It follows immediately that, if a formula of $\cGC^2$ has a finite
model, then it has arbitrarily large finite models, and indeed
infinite models.

As with $\cC^2$, so too with $\cGC^2$, we can limit the nesting of
quantifiers.
\begin{lemma}
Let $\varphi$ be a $\cGC^2$-formula.  There exist $($i$)$ a
quantifier-free $\cGC^2$-formula $\alpha$ with $x$ as its only
variable, $($ii$)$ binary predicates $e_1, \ldots, e_l$, and \linebreak
$f_1, \ldots,
f_m$ $($different from $\approx$$)$, $($iii$)$ quantifier-free $\cGC^2$-formulas
$\beta_1, \ldots, \beta_l$, $($iv$)$ positive integers $C_1, \ldots,
C_m$ with the following property.  If $\varphi^*$ is the $\cGC^2$-formula
\begin{multline}
\forall x \alpha \wedge
\bigwedge_{1 \leq h \leq l} 
   \forall x \forall y (e_h(x,y) \rightarrow (\beta_h \vee x \approx y)) \wedge \\
\bigwedge_{1 \leq i \leq m} 
      \forall x \exists_{=C_i} y (f_i(x,y) \wedge x \not \approx y),
\label{eq:normalformGC2}
\end{multline}
and $C = \max_h C_h$, then $($i$)$ $\varphi^* \models \varphi$, and $($ii$)$
any model of $\varphi$ over a domain having at least $C+1$ elements may
be expanded to a model of $\varphi^*$.
\label{lma:normalformGC2}
\end{lemma}
\begin{proof}
Routine adaptation of standard techniques. See, e.g.~B{\"o}rger {\em
et al.}~\cite{logic:BGG97}, p.~378.
\end{proof}

In view of Lemma~\ref{lma:normalformGC2}, we fix a signature
$\Sigma^*$ of unary and binary predicates and a $\cGC^2$-sentence
$\varphi^*$ over this signature, having the
form~\eqref{eq:normalformGC2}. For the remainder of
Section~\ref{sec:GC2}, all structures will interpret the signature
$\Sigma^*$. We refer to the predicates $f_1, \ldots, f_m$
in~\eqref{eq:normalformGC2} as the {\em counting predicates} of
$\Sigma^*$; and we understand the notions of {\em message type}, {\em
invertible message type}, {\em silent $2$-type} and {\em vacuous}
2-type as in Definition~\ref{def:messagetype}.

For the next definition, if $\pi$ is a 1-type we denote by $\pi[y/x]$
the set of formulas obtained by replacing all occurrences of $x$ in
$\pi$ by $y$. (Recall that 1-types, on our definition, always involve
the variable $x$: so, technically, $\pi[y/x]$ is not a 1-type.)
\begin{definition}
  Let $\pi$ and $\pi'$ be 1-types
  over $\Sigma^*$. Denote by $\pi \times \pi'$ the vacuous 2-type
\begin{equation*}
\pi \cup \pi'[y/x] \cup \{ \neg q(x,y) , \neg q(y,x) \mid \mbox{$q$ a
  binary predicate of $\Sigma^*$} \}.
\end{equation*}
\label{def:vacuous}
\end{definition}
\begin{lemma}
Suppose $\fA \models \varphi^*$, and let $\hat{\fA}$ be the structure
obtained by replacing every silent 2-type in $\fA$ by the
corresponding vacuous 2-type, that is:
\begin{equation*}
{\rm tp}^{\hat{\fA}}[a,b] = 
\begin{cases}
{\rm tp}^\fA[a] \times {\rm tp}^\fA[b] \text{ if ${\rm tp}^\fA[a,b]$ is silent}\\ 
{\rm tp}^\fA[a,b] \text{ otherwise}.
\end{cases}
\end{equation*}
Then $\hat{\fA} \models \varphi^*$. 
\label{lma:silentVacuous}
\end{lemma}
\begin{proof}
Since the 1-types of elements are the same in $\fA$ and $\hat{\fA}$,
$\hat{\fA} \models \forall x \alpha$.  Since the only 2-types
realized in $\hat{\fA}$ but not in $\fA$ are vacuous, and since the
guards in $e_h$ are not satisfied by pairs of elements having vacuous
2-types, $\hat{\fA} \models \bigwedge_{1 \leq h \leq l} \forall x
\forall y (e_h(x,y) \rightarrow (\beta_h \vee x \approx y))$. Since
all elements send the same messages in $\fA$ and $\hat{\fA}$,
$\hat{\fA} \models \bigwedge_{1 \leq i \leq m} \forall x
\exists_{=C_i} y (f_i(x,y) \wedge x \not \approx y)$.
\end{proof}
\begin{lemma}
Suppose that $\fA \models \varphi^*$, and that $B$ and $B'$ are disjoint
subsets of $A$ such that $|B| \geq (mC)^2 +mC +1$, and $|B'| \geq mC
+1$. Then there exist  elements $b \in B$ and $b' \in B'$ such
that ${\rm tp}^\fA[b,b']$ is silent.
\label{lma:bigsets}
\end{lemma}
\begin{proof}
Pick any $B'_0 \subseteq B'$ such that $|B'_0| = mC+1$. Now set
\begin{equation*}
B_0 = \{ b \in B \mid \mbox{for some $b' \in B'_0$, $b'$ sends a message
  to $b$} \}.
\end{equation*} 
Since $\fA \models \varphi^*$, no element of $B'_0$ sends a message to
more than $mC$ other elements, and since $|B'_0| = mC+1$, $|B_0| \leq
mC(mC+1)$.  But $|B| > mC(mC+1)$; so let $b \in B \setminus B_0$.
Again, $b$ can send a message to at most $mC$ elements of $B'_0$, yet
$|B'_0| > mC$; so let $b'$ be an element of $B'_0$ to which $b$ does
not send a message.
\end{proof}

The ensuing analysis hinges on the special notion of a `t-cycle',
which we now proceed to define.  In the sequel, we employ the notions
of {\em path} and {\em cycle} in a graph $G$ in the usual way, where
paths and cycles are not permitted to encounter nodes more than once
(except of course that cycles loop back to their starting points).  We
take the {\em length} of a path $v_0, \ldots, v_l$ to be $l$, and the
{\em length} of a cycle $v_0, \ldots, v_l$ (where $v_l = v_0$) to be
$l$. We insist that, by definition, all cycles have length at least 3.
\begin{definition}
Let $\fA$ be any structure interpreting $\Sigma^*$ over a domain $A$;
let $O \subseteq A$; and let
\begin{align*} 
E = \{ (a,b) \in A^2 \mid \mbox{$a \neq b$ and } &
      \mbox{either ${\rm tp}^\fA[a,b]$ is not vacuous}\\
& \mbox{or $a$ and $b$ are both in $O$} \},
\end{align*}
so that $G= (A, E)$ is a graph. By a {\em t-cycle in} $(\fA,O)$, we
mean a cycle in $G$ containing at least one node lying outside $O$.  A
t-cycle in $(\fA, O)$ is {\em strong} if, for any consecutive
pair of elements $a$ and $b$ in that cycle, either $a$ and $b$ are
both in $O$ or ${\rm tp}^\fA[a,b]$ is an invertible message-type.
\label{def:cycle}
\end{definition}
To motivate these notions, think of $O$ as the set of `observable
elements' of $A$---the elements that will interpret the constants in
some set of ground, function-free literals $\Delta$. By contrast, the
elements of $A \setminus O$ are the `theoretical' elements---elements
whose existence may be perhaps forced by the background theory $\varphi^*$. A t-cycle
is thus a cycle in the graph $G$ of Definition~\ref{def:cycle} which
involves at least one theoretical element.

Our first task is to show that, given any (finite) model $\fA$ of
$\varphi^*$ and any $O \subseteq A$, we can remove all `short' strong
t-cycles in $(\fA,O)$.
\begin{lemma}
Suppose $\fA_0 \models \varphi^*$; and let $O \subseteq A_0$ and $\Omega
> 0$.  We can find a model $\fB \models \varphi^*$ such that: $($i$)$ $O
\subseteq B$; $($ii$)$ $\fA_0 |_O = \fB|_O$; and $($iii$)$ there are
no strong t-cycles in $(\fB,O)$ of length less than
$\Omega$. Moreover, if $\fA_0$ is finite, then we can ensure that
$\fB$ is finite.
\label{lma:bigStrongLoops}
\end{lemma}
\noindent
\begin{proof}
Assume without loss of generality that $\Omega \geq 4$, let
\begin{equation*}
K = 2(|O|+1)((mC)^\Omega - 1)/(mC -1) + 2,
\end{equation*} 
and let $\fA_1, \ldots, \fA_K$ be isomorphic copies of $\fA_0$, with
$A_i \cap A_j = \emptyset$ for all $i,j$ ($0 \leq i < j \leq K$).  Let
$\fA$ (with domain $A$) be the union of $\fA_0$ together with all of
these copies. Formally:
\begin{eqnarray*}
A & = & \bigcup_{0 \leq i \leq K} A_i\\
q^{\fA}  & = & \bigcup_{0 \leq i \leq K} q^{\fA_i} \text{
  for any predicate $q$}.
\end{eqnarray*}
By Lemma~\ref{lma:dup}, $\fA \models \varphi^*$. (Here, we require that
$\varphi^*$ is in $\cGC^2$, not just in $\cC^2$.)  Moreover, if any
element of $A$ sends a message of type $\mu$ in $\fA$, then at least
$K$ elements of $A \setminus O$ do so.

For $a, b \in A$, let us say that $b$ is {\em directly accessible from} $a$ if
either (i) $a = b$, (ii) ${\rm tp}^\fA[a,b]$ is a message-type (not
necessarily invertible), or (iii) $a$ and $b$ are both in $O$;
further, let us say that $b$ {\em is accessible from} $a$ {\em in} $l$
{\em steps}, if there exists a sequence of elements $a_0, \ldots, a_l$
of $A$ such that $a_0 = a$, $a_l = b$ and, for all $i$
($0 \leq i < l$), $a_{i+1}$ is directly accessible from $a_i$. If $a \in A$,
the number of elements accessible from $a$ in $l$ steps is certainly
bounded by $(|O|+1)\sum_{0 \leq i \leq l}(mC)^i$.

Suppose then
\begin{equation*}
\gamma = a_0, a_1, a_2 \ldots,  a_0
\end{equation*}
is a strong t-cycle in $(\fA,O)$ of minimal length $l < \Omega$; and
assume, without loss of generality, that $a_0 \not \in O$.  We modify
$\fA$ (without affecting $\fA|_O$) so as to destroy this t-cycle, taking
care only to create new strong t-cycles of greater length. Let $a =
a_0$ and $b= a_1$, and let $\mu$ be the invertible message-type such
that ${\rm tp}^{\fA}[a,b] = \mu$.

\begin{claim}
There exist pairwise distinct elements $c, d, e, f \in A \setminus O$
such that
\begin{enumerate}
\item ${\rm tp}^\fA[c,d] = \mu$; 
\item neither $c$ nor $d$ is accessible from either $a$ or $b$ in $\Omega - 2$ steps;
\item ${\rm tp}^\fA[e] = {\rm tp}^\fA[a]$, and ${\rm tp}^\fA[f] = {\rm tp}^\fA[b]$; 
\item ${\rm tp}^\fA[e,f]$ is silent;
\item ${\rm tp}^\fA[d,e]$ is not a message-type.
\end{enumerate}
\label{claim:bigConfig}
\end{claim}
\begin{figure}
\begin{center}
\begin{picture}(100,100)(-20,-20)
\put(75,94){$\mu^{-1}$}
\put(18,94){$\mu$}
\put(0,90){\line(1,0){100}}
\put(0,90){\vector(1,0){20}}
\put(100,90){\vector(-1,0){23}}
\put(0,90){\circle*{2}}
\put(100,90){\circle*{2}}
\put(-7,88){$a$}
\put(102,88){$b$}
\put(-10,120){\small \begin{minipage}{4.3cm}
elements accessible from either $a$ or $b$
in $\Omega-2$ steps
\end{minipage}}
\put(50,100){\oval(150,80)}

\put(75,44){$\mu^{-1}$}
\put(18,44){$\mu$}
\put(0,40){\line(1,0){100}}
\put(0,40){\vector(1,0){20}}
\put(100,40){\vector(-1,0){23}}
\put(0,40){\circle*{2}}
\put(100,40){\circle*{2}}
\put(-7,38){$c$}
\put(102,38){$d$}

\put(0,-10){\circle*{2}}
\put(0,-10){\line(2,1){100}}
\put(0,-10){\vector(2,1){80}}
\put(-7,-10){$e$}

\put(100,-10){\circle*{2}}
\put(102,-10){$f$}
\put(0,-10){\line(1,0){100}}
\put(70,26){$($}
\put(80,26){$)$}

\put(0,33){\oval(20,133)}
\put(100,33){\oval(20,133)}
\put(-4,-27){$E$}
\put(96,-27){$F$}
\end{picture}
\end{center}
\caption{The configuration of the claim in the proof of
Lemma~\ref{lma:bigStrongLoops}.  An arrow on a line indicates a
message-type; absence of an arrow on a line indicates a non-message
type; a parenthetical arrow on a line indicates a 2-type which may or
may not be a message-type. For definiteness, $e$ and $f$ have been
drawn outside the set of elements accessible from $a$ or $b$ in
$\Omega-2$ steps; however, this is not required by the claim.}
\label{fig:bigConfig}
\end{figure}
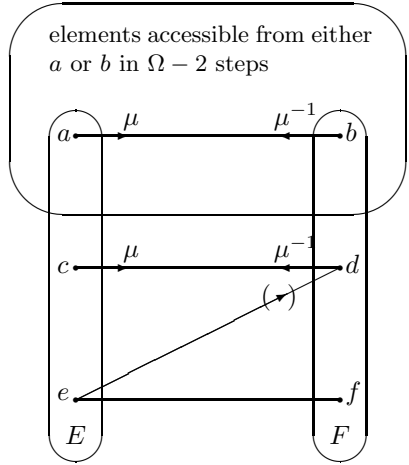
\begin{proof}[Proof of Claim]
Refer to Fig.~\ref{fig:bigConfig}.  The number of elements of $A
\setminus O$ accessible from either $a$ or $b$ in $\Omega-1$ steps is
bounded by
\begin{equation*}
2(|O|+1) \left( \sum_{i=0}^{\Omega -1} (mC)^i \right) = 
      2(|O|+1)({(mC)}^\Omega-1)/((mC)-1) < K.
\end{equation*}
So choose $c \in A \setminus O$ such that $c$ sends a message of type
$\mu$, and $c$ is not accessible from either $a$ or $b$ in $\Omega-1$
steps; and choose $d \in A$ such that ${\rm tp}^\fA[c,d] = \mu$. It
follows that $d$ is not accessible from $a$ or $b$ in $\Omega - 2$
steps. Let $E$ be the set of elements of $A \setminus O$ having the
same 1-type as $a$, and $F$ the set of elements of $A \setminus O$
having the same 1-type as $b$. Now, $E$ and $F$ have cardinality at
least $K$, where, since $\Omega \geq 4$,
\[
K \geq 2((mC)^4 -1)/(mC-1)+2 = 2((mC)^3 + (mC)^2 + mC + 2),
\]
Hence $|E \setminus \{a,b,c,d\}| \geq 2mC((mC)^2 + mC + 1)$; and
similarly, $|F \setminus \{a,b,c,d\}| \geq 2mC((mC)^2 + mC +1)$.
Therefore, we may select subsets $E_1, \ldots, E_{mC}$ of $E \setminus
\{a,b,c,d\}$ and subsets $F'_1, \ldots, F'_{mC}$ of $F \setminus
\{a,b,c,d\}$, each containing at least $(mC)^2 + mC+1$ elements, and
with these $2mC$ sets pairwise disjoint. Applying
Lemma~\ref{lma:bigsets} to $E_i$ and $F_i$ for all $i$ ($1 \leq i \leq
cM$), select $e_i \in E_i$ and $f_i \in F_i$ such that ${\rm
tp}^\fA[e_i,f_i]$ is silent. But $d$ cannot send a message to more
than $mC-1$ of the $e_i$ (since it already sends a message to $c$), so
we may pick $e$ to be some $e_i$ such that ${\rm tp}^\fA[d,e_i]$ is
not a message-type, and $f$ to be the corresponding $f_i$.  The
elements $c$, $d$, $e$ and $f$ then have all the properties required
by the claim.
\end{proof}
Having obtained $c, d, e, f$, and returning to the proof of the lemma,
we modify $\fA$ so as to ensure that the 2-type connecting $a$ and $d$
is silent. (Note that ${\rm tp}^\fA[a,d]$ is certainly not a
message-type, but ${\rm tp}^\fA[d,a]$ might be.) More precisely, we
define the structure $\fA'$ over $A$ to be exactly like $\fA$ except
that
\begin{eqnarray*}
{\rm tp}^{\fA'}[a,d] & = & {\rm tp}^{\fA}[e,f]\\
{\rm tp}^{\fA'}[e,d] & = & {\rm tp}^{\fA}[a,d]\\
{\rm tp}^{\fA'}[e,f] & = & {\rm tp}^{\fA}[e,d].
\end{eqnarray*}
The transformation of $\fA$ into $\fA'$ is depicted in
Fig.~\ref{fig:transform1}.
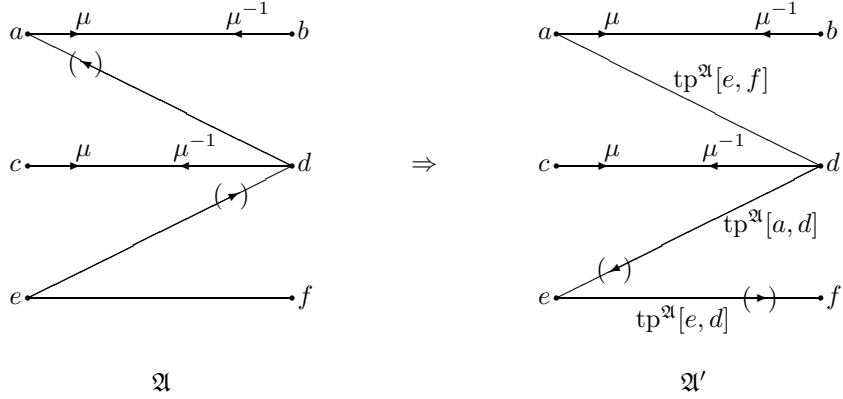
\begin{figure}
\begin{center}
\begin{picture}(350,120)(-20,-30)
\put(75,104){$\mu^{-1}$}
\put(18,104){$\mu$}
\put(0,100){\line(1,0){100}}
\put(0,100){\vector(1,0){20}}
\put(100,100){\vector(-1,0){23}}
\put(0,100){\circle*{2}}
\put(100,100){\circle*{2}}
\put(-7,98){$a$}
\put(102,98){$b$}
\put(0,100){\line(2,-1){100}}
\put(100,50){\vector(-2,1){80}}
\put(15,86){$($}
\put(25,86){$)$}

\put(55,54){$\mu^{-1}$}
\put(18,54){$\mu$}
\put(0,50){\line(1,0){100}}
\put(0,50){\vector(1,0){20}}
\put(100,50){\vector(-1,0){43}}
\put(0,50){\circle*{2}}
\put(100,50){\circle*{2}}
\put(-7,48){$c$}
\put(102,48){$d$}

\put(100,0){\circle*{2}}
\put(0,0){\circle*{2}}
\put(0,0){\line(2,1){100}}
\put(-7,-2){$e$}
\put(102,-2){$f$}
\put(0,0){\vector(2,1){80}}
\put(70,36){$($}
\put(80,36){$)$}
\put(0,0){\line(1,0){100}}
\put(47,-35){$\fA$}

\put(145,48){$\Rightarrow$}

\put(275,104){$\mu^{-1}$}
\put(218,104){$\mu$}
\put(200,100){\line(1,0){100}}
\put(200,100){\vector(1,0){20}}
\put(300,100){\vector(-1,0){23}}
\put(200,100){\circle*{2}}
\put(300,100){\circle*{2}}
\put(193,98){$a$}
\put(302,98){$b$}
\put(300,50){\line(-2,1){100}}
\put(300,50){\vector(-2,-1){80}}
\put(244,80){${\rm tp}^\fA[e,f]$}

\put(255,54){$\mu^{-1}$}
\put(218,54){$\mu$}
\put(200,50){\line(1,0){100}}
\put(200,50){\vector(1,0){20}}
\put(300,50){\vector(-1,0){43}}
\put(200,50){\circle*{2}}
\put(300,50){\circle*{2}}
\put(193,48){$c$}
\put(302,48){$d$}
\put(263,25){${\rm tp}^\fA[a,d]$}

\put(300,0){\circle*{2}}
\put(200,0){\circle*{2}}
\put(200,0){\line(2,1){100}}
\put(193,-2){$e$}
\put(302,-2){$f$}
\put(200,0){\vector(1,0){80}}
\put(215,8){$($}
\put(225,8){$)$}
\put(270,-3){$($}
\put(280,-3){$)$}
\put(230,-11){${\rm tp}^\fA[e,d]$}
\put(200,0){\line(1,0){100}}
\put(247,-35){$\fA'$}
\end{picture}
\end{center}
\caption{Ensuring that ${\rm tp}^{\fA'}[a,d]$ is silent. Types
displayed in the drawing of $\fA'$ are to be read left-to-right: thus,
${\rm tp}^{\fA'}[a,d] = {\rm tp}^\fA[e,f]$, ${\rm tp}^{\fA'}[e,d] =
{\rm tp}^{\fA}[a,d]$, and ${\rm tp}^{\fA'}[e,f] = {\rm
tp}^{\fA}[e,d]$. Lines and arrows are interpreted as in 
Fig.~\ref{fig:bigConfig}.}
\label{fig:transform1}
\end{figure}
The elements $a$, $c$ and $e$ all have the same 1-type in $\fA$;
similarly for $b$, $d$ and $f$. Therefore, these type-assignments are
legitimate, and do not affect the 1-types of any elements, whence
$\fA' \models \forall x \alpha$.  Since no new 2-types are introduced,
$\fA' \models \bigwedge_{1 \leq h \leq l} \forall x \forall y
(e_h(x,y) \rightarrow (\beta_h \vee x \approx y))$. By inspection of
Fig.~\ref{fig:transform1}, every element sends the same messages in
$\fA'$ as in $\fA$ (though to different elements), whence $\fA'
\models \bigwedge_{1 \leq i \leq m} \forall x \exists_{=C_i} y
(f_i(x,y) \wedge x \not \approx y)$. Thus, $\fA' \models \varphi^*$. 
Since $a, e \not \in O$, $\fA'|_O = \fA|_O$; and by construction, ${\rm
tp}^{\fA'}[a,d]$ is silent. Note also that $\fA$ and $\fA'$ never
differ with respect to any invertible message-types: in particular,
the strong t-cycles in $(\fA,O)$ are exactly the strong t-cycles in
$(\fA',O)$.

We are now ready to destroy the strong t-cycle $\gamma$ in $(\fA',
O)$.  Let $\fA''$ be exactly like $\fA'$, except that
\begin{align*}
{\rm tp}^{\fA''}[a,b] = & {\rm tp}^{\fA'}[a,d] & 
{\rm tp}^{\fA''}[a,d] = & {\rm tp}^{\fA'}[a,b]\\
{\rm tp}^{\fA''}[c,b] = & {\rm tp}^{\fA'}[c,d] &
{\rm tp}^{\fA''}[c,d] = & {\rm tp}^{\fA'}[c,b].
\end{align*}
The transformation of $\fA'$ into $\fA''$ is depicted in
Fig.~\ref{fig:transform2}.
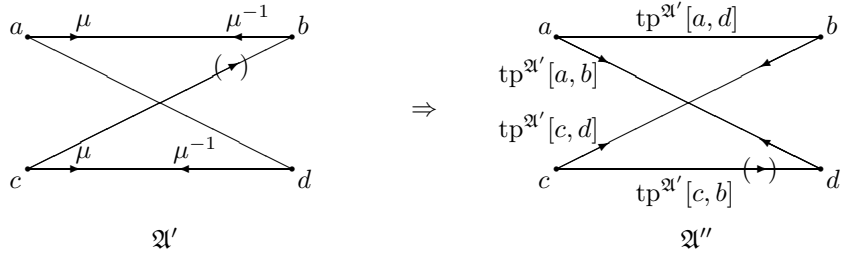
\begin{figure}
\begin{center}
\begin{picture}(350,100)(-20,30)
\put(75,104){$\mu^{-1}$}
\put(18,104){$\mu$}
\put(0,100){\line(1,0){100}}
\put(0,100){\vector(1,0){20}}
\put(100,100){\vector(-1,0){23}}
\put(0,100){\circle*{2}}
\put(100,100){\circle*{2}}
\put(-7,101){$a$}
\put(102,101){$b$}
\put(100,50){\line(-2,1){100}}
\put(0,50){\line(2,1){100}}
\put(0,50){\vector(2,1){80}}
\put(70,86){$($}
\put(80,86){$)$}

\put(55,54){$\mu^{-1}$}
\put(18,54){$\mu$}
\put(0,50){\line(1,0){100}}
\put(0,50){\vector(1,0){20}}
\put(100,50){\vector(-1,0){43}}
\put(0,50){\circle*{2}}
\put(100,50){\circle*{2}}
\put(-7,43){$c$}
\put(102,43){$d$}
\put(47,20){$\fA'$}

\put(145,70){$\Rightarrow$}

\put(200,100){\line(1,0){100}}
\put(200,100){\line(2,-1){100}}
\put(200,100){\vector(2,-1){20}}
\put(270,47){$($}
\put(280,47){$)$}
\put(300,50){\vector(-2,1){23}}
\put(200,100){\circle*{2}}
\put(300,100){\circle*{2}}
\put(193,101){$a$}
\put(302,101){$b$}
\put(300,50){\line(-2,1){100}}
\put(178,83){${\rm tp}^{\fA'}[a,b]$}
\put(178,62){${\rm tp}^{\fA'}[c,d]$}
\put(230,104){${\rm tp}^{\fA'}[a,d]$}

\put(200,50){\line(1,0){100}}
\put(200,50){\vector(1,0){80}}
\put(200,50){\line(2,1){100}}
\put(200,50){\vector(2,1){20}}
\put(300,100){\vector(-2,-1){23}}
\put(200,50){\circle*{2}}
\put(300,50){\circle*{2}}
\put(193,43){$c$}
\put(302,43){$d$}
\put(230,38){${\rm tp}^{\fA'}[c,b]$}
\put(247,20){$\fA''$}
\end{picture}
\end{center}
\caption{Destroying a strong t-cycle: the two-types in $\fA''$ are to be read
  from left to right; thus, ${\rm tp}^{\fA''}[a,b] = {\rm
  tp}^{\fA'}[a,d]$, ${\rm tp}^{\fA''}[a,d] = {\rm tp}^{\fA'}[a,b]$,
  ${\rm tp}^{\fA''}[c,b] = {\rm tp}^{\fA'}[c,d]$ and ${\rm
  tp}^{\fA''}[c,d] = {\rm tp}^{\fA'}[c,b]$. Lines and arrows are
  interpreted as in Fig.~\ref{fig:bigConfig}.}
\label{fig:transform2}
\end{figure}
Again, these assignments are legitimate, with 1-types unaffected; no
new 2-types are introduced; and every element of $A$ sends the same
messages in $\fA''$ as it does in $\fA'$ (though to different
elements). Thus $\fA'' \models \varphi^*$. Since $a, c \not \in O$,
$\fA''|_O = \fA'|_O = \fA|_O$; and by construction, $\gamma$ is not a
strong t-cycle in $(\fA'',O)$. Moreover, we claim that any sequence
$\gamma'$ which is a strong t-cycle in $(\fA'', O)$, but not in
$(\fA', O)$, is longer than $\gamma$. To show this, we suppose
$|\gamma'| \leq |\gamma| < \Omega$, and derive a contradiction.  Since
$\gamma'$ is not a strong t-cycle in $(\fA',O)$, at least one of the
pairs $(a,d)$, $(d,a)$, $(b,c)$ or $(c,b)$ is consecutive in
$\gamma'$; so suppose, without loss of generality, that $(a,d)$
is. Indeed, by starting the cycle $\gamma'$ at $d$, we may write
\begin{equation*}
\gamma' = d, \ldots, a, d.
\end{equation*}
Now $b$ certainly occurs in $\gamma'$. For otherwise, all consecutive
pairs of $\gamma'$ except $(a,d)$ send each other messages in $\fA'$,
contradicting the fact that $d$ is not accessible from $a$ in
$\Omega-2$ steps. In fact, an exactly similar argument shows that $(c,
b)$ occurs as a consecutive pair in $\gamma'$, since $d$ is not
accessible from $b$ in $\Omega-2$ steps either. Thus, we may write:
\begin{equation*}
\gamma' = d, c_1, \ldots, c_s, c,b, b_1, \ldots, b_t,a,d,
\end{equation*}
($s,t \geq 0$). Returning to the structure $\fA'$, then, we see that
\begin{align*}
\gamma_1 &= d, c_1, \ldots, c_s, c, d\\
\gamma_2 &= b, b_1, \ldots, b_t, a, b
\end{align*}
are strong t-cycles in $(\fA', O)$; and so, by the minimality of
$\gamma$ in $\fA'$, we have $s+2 \geq |\gamma|$ and $t+2 \geq
|\gamma|$.  It follows that $|\gamma'| = s+t+4 \geq 2|\gamma| >
|\gamma|$, a contradiction.

Thus, in transforming $\fA$ into $\fA''$, we destroy one strong
t-cycle of length less than $\Omega$, and create only longer strong
t-cycles.  Proceeding in this way, then, we eventually destroy all
strong t-cycles of length less than $\Omega$.
\end{proof}

Our next task is to show that, given any (finite) model $\fA$ of
$\varphi^*$ and any $O \subseteq A$, we can remove all `short' t-cycles
in $(\fA,O)$, strong or otherwise.
\begin{lemma}
Suppose $\fA_0 \models \varphi^*$; and let $O \subseteq A_0$ and $\Omega
> 0$.  We can find a model $\fB \models \varphi^*$ such that: $($i$)$ $O
\subseteq B$; $($ii$)$ $\fA_0 |_O = \fB|_O$; and $($iii$)$ there are
no t-cycles in $(\fB,O)$ of length less than $\Omega$. Moreover, if
$\fA_0$ is finite, then we can ensure that $\fB$ is finite.
\label{lma:bigLoops}
\end{lemma}
\begin{proof}
By Lemma~\ref{lma:bigStrongLoops}, let $\fA$ be a finite or countable
model of $\varphi^*$, with $\fA$ finite if $\fA_0$ is, such that: (i) $O
\subseteq A$; (ii) $\fA_0|_O = \fA|_O$; and (iii) there are no strong
t-cycles in $(\fA, O)$ of length less than $\Omega$.  Let
\begin{equation*}
S = \{ \langle a,b \rangle \in A^2 \mid 
  \mbox{$a \neq b$ and ${\rm tp}^\fA[a,b]$ is a non-invertible message-type}\},
\end{equation*}
and let $Y = |S|$. Obviously, if $\fA$ is finite, then so is $Y$.  In
addition, let $S^{*\Omega}$ be the set of sequences of elements of $S$
of length $\leq \Omega$. We denote the length of $\sigma \in
S^{*\Omega}$ by $|\sigma|$; we write empty sequence as $\epsilon$ and
the concatenation of sequences $\sigma$ and $\tau$ as $\sigma \tau$;
as usual, we identify sequences of length 1 with the corresponding
elements of $S$.

Let $\fA_\epsilon = \fA$.  For $\sigma \in S^{*\Omega} \setminus
\{\epsilon\}$, let $\fA_\sigma$ be a new copy of $\fA$, with domain
$A_\sigma$; and for any $a \in A$, denote by $a_\sigma$ the
corresponding element of $A_\sigma$. We assume that the $A_\sigma$
($\sigma \in S^{*\Omega}$) are pairwise disjoint.  Now let $\fA^*$ be
given by:
\begin{eqnarray*}
A^* & = & \bigcup_{\sigma \in S^{*\Omega}} A_\sigma\\
q^{\fA^*}  & = & \bigcup_{\sigma \in S^{*\Omega}} q^{\fA_\sigma} \text{
  for any predicate $q$}.
\end{eqnarray*}
Note that $O \subseteq A \subseteq A^*$.  We may picture $\fA^*$ as a
tree of copies of $\fA$, with $\fA_\epsilon = \fA$ at the root,
and having branching factor $Y$. We notionally divide the tree into
tiers, taking the root to be the first tier, and the leaves to be the
$(\Omega+1)$th tier.  The case where $Y$ is finite is illustrated in
Fig.~\ref{fig:tree}; the case where $Y = \aleph_0$ may be pictured
analogously.
\begin{figure}
\begin{center}
\begin{picture}(225,100)
\put(-15,40){\vector(0,-1){37}}
\put(-15,40){\vector(0,1){78}}
\put(-40,58){$\Omega+1$}
\put(113,114){$\fA_\epsilon = \fA$}
\put(115,110){\circle*{4}}
\put(115,110){\line(-2,-1){70}}
\put(115,110){\line(2,-1){70}}
\put(45,75){\circle*{4}}
\put(185,75){\circle*{4}}
\put(45,75){\line(-1,-1){25}}
\put(45,75){\line(1,-1){25}}
\put(185,75){\line(-1,-1){25}}
\put(185,75){\line(1,-1){25}}
\put(28,79){$\fA_{s_1}$}
\put(183,79){$\fA_{s_Y}$}
\put(113,75){$\ldots$}
\put(115,82){\vector(-1,0){45}}
\put(115,82){\vector(1,0){45}}
\put(113,85){$Y$}
\put(20,50){\circle*{4}}
\put(70,50){\circle*{4}}
\put(160,50){\circle*{4}}
\put(210,50){\circle*{4}}
\put(20,50){\line(-1,-3){15}}
\put(70,50){\line(1,-3){15}}
\put(160,50){\line(-1,-3){15}}
\put(210,50){\line(1,-3){15}}
\put(21,40){$\fA_{s_1,s_1}$}
\put(74,47){$\fA_{s_1,s_Y}$}
\put(161,40){$\fA_{s_Y,s_1}$}
\put(214,47){$\fA_{s_Y,s_Y}$}
\put(45,54){\vector(1,0){17}}
\put(45,54){\vector(-1,0){17}}
\put(42,56){$Y$}
\put(185,54){\vector(1,0){17}}
\put(185,54){\vector(-1,0){17}}
\put(182,56){$Y$}
\put(5,5){\circle*{4}}
\put(85,5){\circle*{4}}
\put(145,5){\circle*{4}}
\put(225,5){\circle*{4}}
\put(5,5){\line(1,0){80}}
\put(145,5){\line(1,0){80}}
\put(2,-5){$\fA_{s_1, s_1, \ldots, s_1}$}
\put(83,-5){$\fA_{s_1,s_Y, \ldots, s_Y}$}
\put(142,-5){$\fA_{s_Y,s_1, \ldots s_1}$}
\put(223,-5){$\fA_{s_Y,s_Y, \ldots, s_Y}$}
\end{picture}
\end{center}
\caption{Organization of $\fA^*$ as a tree of copies of $\fA_0$, in the
case where $Y = |S|$ is finite; for legibility, the elements of $S$
are numbered, arbitrarily, as $s_1, \ldots s_Y$.}
\label{fig:tree}
\end{figure}
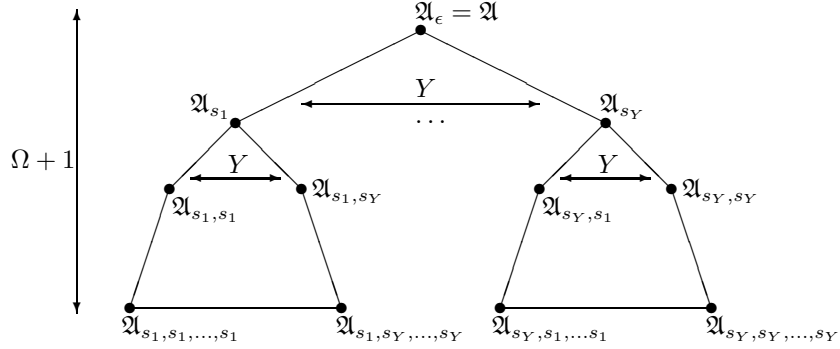
By Lemma~\ref{lma:dup}, $\fA^* \models \varphi^*$.  (Here, we require
that $\varphi^*$ is in $\cGC^2$, not just in $\cC^2$.) Moreover, there
are no strong t-cycles in $(\fA^*,O)$ of length less than $\Omega$.

We modify $\fA^*$ as follows to obtain a structure $\fB$ over the
domain $B = A^*$.  As a first (easy) step, if $a$ and $b$ are any
distinct elements of $A^*$, not both in $O$, such that ${\rm
tp}^\fA[a,b]$ is silent but not vacuous, we can apply
Lemma~\ref{lma:silentVacuous}, and replace ${\rm tp}^\fA[a,b]$ with
the vacuous 2-type ${\rm tp}^\fA[a] \times {\rm tp}^\fA[b]$. (Notice
that this transformation does not affect $\fA^*|_O$.) Hence, we may
assume that, if $(a,b)$ is a consecutive pair in some t-cycle in
$(\fA^*,O)$, with $a, b$ not both in $O$, then at least one of ${\rm
tp}^\fA[a,b]$ and ${\rm tp}^\fA[b,a]$ is a message-type. Furthermore,
since there are no strong t-cycles in $(\fA^*,O)$ of length less than
$\Omega$, any t-cycle in $(\fA^*,O)$ of length less than $\Omega$
contains at least one consecutive pair $(a,b)$, such that: (i) $a$ and
$b$ are not both in $O$, and (ii) {\em exactly} one of ${\rm
tp}^\fA[a,b]$ and ${\rm tp}^\fA[b,a]$ is a message-type (and hence a
non-invertible message-type).

We obtain $\fB$ from $\fA^*$ by re-directing non-invertible messages
in successive tiers of the tree in Fig.~\ref{fig:tree} as
follows. First, we consider the structure $\fA_\epsilon = \fA$ at the
root of the tree. Let $a, b$ be any distinct elements of $A$, not both
in $O$. If ${\rm tp}^{\fA}[a,b]$ is a non-invertible message-type
$\mu$, then we divert the message which $a$ sends to $b$ in $\fA^*$ so
that it instead points to the element corresponding to $b$ in the
structure at the $\langle a,b \rangle$th position in the second tier
of the tree in Fig.~\ref{fig:tree}. Formally, we set
\begin{eqnarray*}
{\rm tp}^{\fB}[a,b] & = & {\rm tp}^{\fA^*}[a] \times
                                        {\rm tp}^{\fA^*}[b] \\
{\rm tp}^{\fB}[a,b_{\langle a,b\rangle}] & = & 
                                        {\rm tp}^{\fA^*}[a,b].
\end{eqnarray*}
Otherwise, we leave the elements of $\fA_\epsilon$ unaffected.
This transformation is depicted in Fig.~\ref{fig:redirection}.

Next, we consider the copies of $\fA$ in tiers 2 to $\Omega$: i.e.
those $\fA_\sigma$ such that $1 \leq |\sigma| < \Omega$.  Let $a, b$
be any distinct elements of $A$. If ${\rm tp}^{\fA}[a,b]$ is a
non-invertible message-type $\mu$, then we divert the message which
$a_\sigma$ sends to $b_\sigma$ in $\fA^*$ so that it instead points to
the element corresponding to $b$ in the copy of $\fA$ located at the
$\langle a,b \rangle$th daughter of $\fA_\sigma$. Formally, we set
\begin{eqnarray*}
{\rm tp}^{\fB}[a_\sigma,b_\sigma] & = & {\rm tp}^{\fA^*}[a_\sigma] \times
                                        {\rm tp}^{\fA^*}[b_\sigma] \\
{\rm tp}^{\fB}[a_\sigma,b_{\sigma\langle a,b\rangle}] & = & 
                                        {\rm tp}^{\fA^*}[a_\sigma,b_\sigma].
\end{eqnarray*}
Otherwise, we leave the elements of $\fA_\sigma$ unaffected.

Finally, we consider the copies of $\fA$ in the bottom tier: i.e.
those $\fA_\sigma$ such that $|\sigma| = \Omega$.  Let $a, b$ be any
distinct elements of $A$. If ${\rm tp}^{\fA}[a,b]$ is a non-invertible
message-type $\mu$, then we divert the message which $a_\sigma$ sends
to $b_\sigma$ in $\fA^*$ so that it instead loops back to the element
corresponding to $b$ in the structure located at the $\langle a,b
\rangle$th node of the {\em second} tier of the tree. Formally, we set
\begin{eqnarray*}
{\rm tp}^{\fB}[a_\sigma,b_\sigma] & = & {\rm tp}^{\fA^*}[a_\sigma] \times
                                        {\rm tp}^{\fA^*}[b_\sigma] \\
{\rm tp}^{\fB}[a_\sigma,b_{\langle a,b\rangle}] & = & 
                                        {\rm tp}^{\fA^*}[a_\sigma,b_\sigma].
\end{eqnarray*}
Otherwise, we leave the elements of $\fA_\sigma$ unaffected.
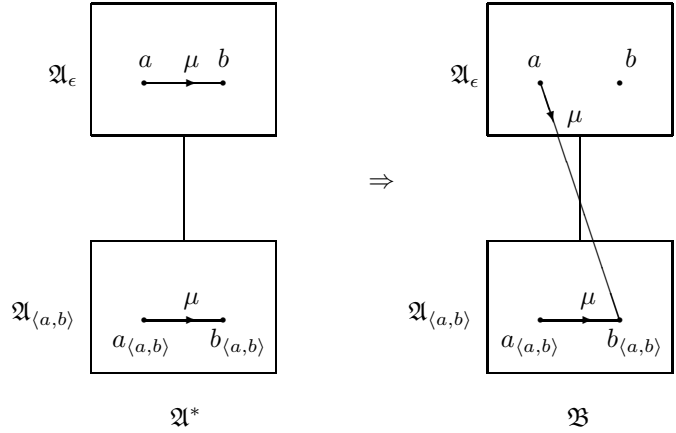
\begin{figure}
\begin{center}
\begin{picture}(100,100)(-20,-20)
\put(-90,20){$\fA_{\langle a,b \rangle}$}
\put(-60,0){\line(1,0){70}}
\put(-60,0){\line(0,1){50}}
\put(10,50){\line(-1,0){70}}
\put(10,50){\line(0,-1){50}}
\put(-40,20){\circle*{2}}
\put(-10,20){\circle*{2}}
\put(-40,20){\line(1,0){30}}
\put(-40,20){\vector(1,0){20}}
\put(-52,10){$a_{\langle a, b \rangle}$}
\put(-15,10){$b_{\langle a, b \rangle}$}
\put(-25,26){$\mu$}

\put(-30,-20){$\fA^*$}

\put(60,20){$\fA_{\langle a,b \rangle}$}
\put(90,0){\line(1,0){70}}
\put(90,0){\line(0,1){50}}
\put(160,50){\line(-1,0){70}}
\put(160,50){\line(0,-1){50}}
\put(110,20){\circle*{2}}
\put(140,20){\circle*{2}}
\put(110,20){\line(1,0){30}}
\put(110,20){\vector(1,0){20}}
\put(95,10){$a_{\langle a, b \rangle}$}
\put(135,10){$b_{\langle a, b \rangle}$}
\put(125,26){$\mu$}
\put(120,-20){$\fB$}

\put(45,70){$\Rightarrow$}

\put(-76,110){$\fA_\epsilon$}
\put(-60,90){\line(1,0){70}}
\put(-60,90){\line(0,1){50}}
\put(10,140){\line(-1,0){70}}
\put(10,140){\line(0,-1){50}}
\put(-25,90){\line(0,-1){40}}
\put(-40,110){\circle*{2}}
\put(-10,110){\circle*{2}}
\put(-40,110){\line(1,0){30}}
\put(-40,110){\vector(1,0){20}}
\put(-42,116){$a$}
\put(-12,116){$b$}

\put(-25,116){$\mu$}

\put(76,110){$\fA_\epsilon$}
\put(90,90){\line(1,0){70}}
\put(90,90){\line(0,1){50}}
\put(160,140){\line(-1,0){70}}
\put(160,140){\line(0,-1){50}}
\put(125,90){\line(0,-1){40}}
\put(110,110){\circle*{2}}
\put(140,110){\circle*{2}}
\put(110,110){\line(1,-3){30}}
\put(110,110){\vector(1,-3){5}}
\put(105,116){$a$}
\put(142,116){$b$}
\put(120,95){$\mu$}
\end{picture}
\end{center}
\caption{Re-direction of non-invertible messages in $\fA_\epsilon$ in
the proof of Lemma~\ref{lma:bigLoops}.}
\label{fig:redirection}
\end{figure}

It is obvious that these assignments are legitimate, leave 1-types
unaffected, introduce no new 2-types, and leave the number of messages
of each type sent by any element unaffected. Hence, $\fB \models
\varphi^*$. It is equally obvious that $\fB|_O = \fA^*|_O = \fA_0|_O$,
and that there are no t-cycles in $(\fB, O)$ of length less than
$\Omega$.
\end{proof}
We remark that the method of removing short t-cycles used in
Lemma~\ref{lma:bigLoops} works only for cycles featuring
non-invertible message types. In particular, the large `fan-in' at
elements of structures in the second tier requires that the
message-types being redirected are non-invertible.

\section{Data-complexity of query-answering and \newline
finite query-answering}
\label{sec:dataGC2}
In this section, we prove that the query-answering and finite
query-answering problems with respect to a positive conjunctive query
$\psi(\bar{y})$ and a formula $\varphi$ of $\cGC^2$ are in the class
co-NP. Lemma~\ref{lma:bigLoops} plays a key role in this proof, by
allowing us to re-write positive conjunctive queries as disjunctions
of queries involving only two variables (at which point we can apply
Theorem~\ref{theo:dataSatC2}). The remainder of the proof is largely a
matter of book-keeping.

We begin with a generalization of the observation that $\forall x
\forall y \theta(x,y)$ is logically equivalent to $\forall x
\theta(x,x) \wedge \forall x \forall y (x \not \approx y \rightarrow
\theta(x,y))$. We employ the following notation.  Fix some set of
constants $K$ and tuple of variables $\bar{x} = x_1, \ldots, x_n$.
Let $\Xi$ be the set of all functions $\xi: \bar{x} \rightarrow
\bar{x} \cup K$.  For each $\xi \in \Xi$, denote by $\bar{x}_\xi$ the
(possibly empty) tuple of variables $\xi(x_1), \ldots, \xi(x_n)$ with
all constants and duplicates removed. Further, for any formula
$\theta$, denote by $\theta_\xi$ the result of simultaneously
substituting the terms $\xi(x_1)$, \ldots, $\xi(x_n)$ for all free
occurrences of the respective variables $x_1$, \ldots, $x_n$ in
$\theta$.
\begin{lemma}
Let $\bar{x}$ be a tuple of variables, $K$ a finite set of constants, and
$\Xi$ the set of all functions $\xi: \bar{x} \rightarrow \bar{x} \cup
K$.  If $\theta$ is any formula, then $\forall \bar{x} \theta$ is
logically equivalent to
\begin{equation}
\bigwedge_{\xi \in \Xi} \forall \bar{x}_\xi 
       \big( \big(
        \bigwedge_{\substack{x \in \bar{x}_\xi\\
                             c  \in K}} 
                             x \not \approx c
        \wedge
        \bigwedge_{\substack{x, x'  \in \bar{x}_\xi\\
                             x  \neq x'}} 
               x \not \approx x' \big)  \rightarrow \theta_\xi \big). 
\label{eq:simpleConjunctionOverAssignments}
\end{equation}
In fact, let $\Xi_1$ and $\Xi_2$ be disjoint $($possibly empty$)$ subsets
of $\Xi$ such that $\Xi_1 \cup \Xi_2 = \Xi$. Then $\forall \bar{x}
\theta$ is logically equivalent to
\begin{equation}
\left\{
\bigwedge_{\xi \in \Xi_1} \forall \bar{x}_\xi 
       \big( \big(
        \bigwedge_{\substack{x \in \bar{x}_\xi\\
                             c  \in K}} 
                             x \not \approx c
        \wedge
        \bigwedge_{\substack{x, x'  \in \bar{x}_\xi\\
                             x  \neq x'}} 
               x \not \approx x' \rightarrow \theta_\xi \big) \big)
	       \right\}
\wedge
\left\{
\bigwedge_{\xi \in \Xi_2} \forall \bar{x}_\xi 
       \theta_\xi
\right\}. 
\label{eq:conjunctionOverAssignments}
\end{equation}
\label{lma:conjunctionOverAssignments}
\end{lemma}
\begin{proof}
Denote by $\varphi_1$ the
formula~\eqref{eq:simpleConjunctionOverAssignments}, and by $\varphi_2$
the formula~\eqref{eq:conjunctionOverAssignments}. It is obvious that
$\models \forall \bar{x} \theta \rightarrow \varphi_2$, $\models \varphi_2
\rightarrow \varphi_1$, and $\models  \varphi_1 \rightarrow \forall \bar{x} \theta$.
\end{proof}

The next lemma allows us to remove individual constants from universally
quantified formulas at the expense of adding some ground literals.
\begin{lemma}
Let $\varphi$ be a formula, $\Theta$ a set of formulas, and $c$ an
individual constant. Let $p$ be a new unary predicate and $z$ a new
variable $($`new' means `not occurring in $\varphi$ or $\Theta$'$)$.
Denote by $\varphi'$ the result of replacing all occurrences of $c$ in
$\varphi$ by $z$, and let $\psi$ be the formula $\forall z(\varphi' \vee
\neg p(z))$.  Then the sets of formulas $\Theta \cup \{ \varphi \}$ and
$\Theta \cup \{p_c(c), \psi \}$ are satisfiable over the same domains.
\label{lma:getRidOfConstants}
\end{lemma}
\begin{proof}
Obviously, $\{p_c(c), \psi \} \models \varphi$. On the other hand, if
$\fA \models \Theta \cup \{ \varphi \}$, expand $\fA$ to a structure
$\fA'$ by setting $p^{\fA'} = \{c^\fA\}$.
\end{proof}

Recall that a {\em clause} is a disjunction of literals (with the {\em
empty clause}, $\bot$, allowed), and that a clause is {\em negative}
if all its literals are negative. In the sequel, we continue to
confine attention to signatures involving only unary and binary
predicates together with individual constants.
\begin{definition}
Let $\eta$ be a clause, let $T$ be the set of terms (variables or
constants) occurring in $\eta$, and let
\begin{align*} 
E = \{ (t_1,t_2) \in T^2 \mid \mbox{$t_1 \neq t_2$ and } &
      \mbox{either $t_1, t_2$ both occur in some literal of $\eta$}\\
& \mbox{or $t_1$ and $t_2$ are both constants} \}.
\end{align*}
Denote the graph $(T, E)$ by $G_\eta$. (We allow the empty graph for
the case $\eta = \bot$.)  We say $\eta$ is {\em v-cyclic} if $G_\eta$
contains a cycle (in the usual graph-theoretic sense) at least one of
whose nodes is a variable; otherwise, we say $\eta$ is {\em
v-acyclic}.
\label{def:cyclic}
\end{definition}
\begin{definition}
Let $K$ be a set of individual constants.  A {\em
v-formula} ({\em with respect to $K$}) is a sentence of the form
\begin{equation}
\forall \bar{x} \big( \big(
        \bigwedge_{\substack{x \in \bar{x}\\
                             c  \in K}} 
                             x \not \approx c
        \wedge
        \bigwedge_{\substack{x, x'  \in \bar{x}\\
                             x  \neq x'}} 
               x \not \approx x' \big) \rightarrow \eta \big),
\label{eq:v-formula}
\end{equation}
where $\eta$ is a v-cyclic negative clause.
\label{def:v-formulas}
\end{definition}
The intuition behind v-formulas is that they provide a counterpart to
the notion of a t-cycle in a pair $(\fA,O)$, given in
Definition~\ref{def:cycle}. Specifically:
\begin{remark}
Let $\fA$ be a structure, $K$ the set of individual constants
interpreted by $\fA$, and $O = \{c^\fA \mid c \in K \}$. Suppose that
distinct individual constants in $K$ have distinct interpretations in
$\fA$. Let $\upsilon$ be a v-formula with respect to $K$.  If $\fA
\not \models \upsilon$, then there is a t-cycle in $(\fA,O)$ of length
at most $\lVert \upsilon \rVert$.
\label{remark:cyclic}
\end{remark}
\begin{definition}
Let $\eta$ be a clause. We call $\eta$ {\em splittable} if, by
re-ordering its literals, it can be written as $\eta_1 \vee \eta_2$,
where ${\rm Vars}(\eta_1) \cap {\rm Vars}(\eta_2) = \emptyset$;
otherwise, $\eta$ is {\em unsplittable}.
\label{def:splittable}
\end{definition}
\begin{remark}
Let $\eta$ be a non-ground clause. If $\eta$ is unsplittable
and v-acyclic, then it contains at most one individual constant.
\label{remark:unsplittable}
\end{remark}
\begin{lemma}
Let $\eta(x,\bar{x})$ be a negative clause with no individual
constants, involving exactly the variables $x,\bar{x}$.  Suppose
further that $\eta(x,\bar{x})$ is non-empty, unsplittable and
v-acyclic.  Then there exists a $\cGC^2$-formula of $\psi(x)$ such
that $\forall \bar{x} \eta(x,\bar{x})$ and $\psi(x)$ are logically
equivalent.
\label{lma:pureTformulas}
\end{lemma}
\begin{proof}
We proceed by induction on the number of variables involved. If
$\bar{x}$ is the empty tuple, there is nothing to prove, so suppose
otherwise.  Since $\eta$ is unsplittable and v-acyclic, and contains
the variable $x$, $G_\eta$ may be viewed as a tree with $x$ at the
root. Let $x_1, \ldots, x_n$ be the immediate descendants of $x$ in
the tree $G_\eta$. Further, for all $i$ ($1 \leq i \leq n$), let
$\bar{x}_i$ be a (possibly empty) tuple consisting of those variables
in $\bar{x}$ which are proper descendants of $x_i$ in $G_\eta$.  Then
$\forall \bar{x} \eta(x,\bar{x})$ is logically equivalent to some
formula
\begin{equation*}
\delta(x) \vee 
  \bigvee_{1 \leq i \leq n} \forall x_i  (\epsilon_i(x,x_i) \vee 
                                  \forall \bar{x}_i \eta_i(x_i,\bar{x}_i)),
\end{equation*}
where $\delta(x)$ is a negative clause involving exactly the variables
$\{x\}$, and, for all $i$ ($1 \leq i \leq n$): (i) $\epsilon_i(x,x_i)$
is a non-empty negative clause each of whose literals involves the
variables $\{x, x_i\}$, and (ii) $\eta_i(x_i,\bar{x}_i)$ is a negative
clause which involves exactly the variables $\{x_i\} \cup \bar{x}_i$.
By inductive hypothesis, there exists a $\cGC^2$-formula $\psi_i(x_i)$
logically equivalent to $\forall \bar{x}_i \eta_i(x_i,\bar{x}_i)$.  But then $\forall
\bar{x} \eta(x,\bar{x})$ is logically equivalent to
\begin{equation*}
\delta(x) \vee 
  \bigvee_{1 \leq i \leq n} \forall y (\epsilon_i(x,y) \vee \psi_i(y)),
\end{equation*}
which in turn is trivially logically equivalent to a $\cGC^2$-formula.
\end{proof}
\begin{lemma}
Let $\varphi$ be a $\cGC^2$-formula, $\Delta$ a finite set of ground,
function-free literals, and $\Upsilon$ a finite set of v-formulas.
Suppose that $\Delta$ contains the literal $c \not \approx d$ for all
distinct individual constants $c$, $d$ occurring in $\Delta \cup
\Upsilon$. Then $\Delta \cup \{\varphi \} \cup \Upsilon$ is
$($finitely$)$ satisfiable if and only if $\Delta \cup \{\varphi\}$ is
$($finitely$)$ satisfiable.
\label{lma:useBigLoops}
\end{lemma}
\begin{proof}
The only-if direction is trivial.  So suppose $\fA_0^+$ is a (finite)
model of $\{ \varphi \} \cup \Delta$, with domain $A_0$. Let $O \subseteq
A_0$ be the set of elements interpreting the individual constants in
$\Delta \cup \Upsilon$, and let $\fA_0$ be the reduct of $\fA_0^+$
obtained by ignoring the interpretations of those individual
constants.

Let $\varphi^*$ and $C$ be obtained from $\varphi$ as in
Lemma~\ref{lma:normalformGC2}. Let $\fA_1, \ldots, \fA_C$ be
isomorphic copies of $\fA_0$ with the domains $A_i$ ($0 \leq i \leq
C$) pairwise disjoint; and let $\fA$ be the union of these models as
in Lemma~\ref{lma:dup}.  Thus, $O \subseteq A_0 \subseteq A$, $\fA
\models \varphi$, and $|A| > C$.  By Lemma~\ref{lma:normalformGC2}, let
$\fA'$ be an expansion of $\fA$ such that $\fA' \models \varphi^*$.
Obviously, $\fA'$ is finite if $\fA_0^+$ is.

Let $\Omega > \lVert \upsilon \rVert$ for all $\upsilon \in \Upsilon$.
Applying Lemma~\ref{lma:bigLoops} to $\fA'$, let $\fB$ be a model of
$\varphi^*$ (and hence of $\varphi$), finite if $\fA'$ is finite , such
that: (i) $O \subseteq B$; (ii) $\fB|_O = \fA'|_O = \fA_0|_O$; and
(iii) there are no t-cycles in $(\fB, O)$ of length less than
$\Omega$. Let $\fB^+$ be the expansion of $\fB$ obtained by
interpreting any constants as in $\fA_0^+$. Thus, $\fB^+ \models
\Delta \cup \{ \varphi \}$.  If $\fB^+$ fails to satisfy some formula in
$\Upsilon$ of the form~\eqref{eq:v-formula}, then, by
Remark~\ref{remark:cyclic}, there is a t-cycle in $(\fB, O)$ of length
less than $\Omega$, which is impossible. Hence $\fB^+ \models
\Upsilon$, as required.
\end{proof}
\begin{theorem}
For any $\cGC^2$-sentence $\varphi$ and any positive conjunctive query
$\psi(\bar{y})$, both $\cQ_{\varphi, \psi(\bar{y})}$ and $\cFQ_{\varphi,
\psi(\bar{y})}$ are in {\rm co-NP}.
\label{theo:queryAnswering}
\end{theorem}
\begin{proof}
We give the proof for $\cFQ_{\varphi, \psi(\bar{y})}$; the proof for
  $\cQ_{\varphi, q(\bar{y})}$ is analogous.

Let an instance $\langle \Delta, \bar{a} \rangle$ of $\cFQ_{\varphi,
\psi(\bar{y})}$ be given, where $\Delta$ is a set of ground,
function-free literals, and $\bar{a}$ a tuple of individual constants.
By re-naming individual constants if necessary, we may assume that the
constants $\bar{a}$ all have codes of fixed length, so that $\bar{a}$
may be regarded as a constant.  Let $n = \lVert \Delta \rVert$,
then. The instance $\langle \Delta, \bar{a} \rangle$ is positive if
and only if $\psi(\bar{a})$ is true in every finite model of $\Delta
\cup \{\varphi \}$.  Hence, it suffices to give a non-deterministic
procedure for determining the finite satisfiability of the formula
\begin{equation}
\bigwedge \Delta \wedge \varphi \wedge \neg \psi(\bar{a}),
\label{eq:coProblem}
\end{equation}
running in time bounded by a polynomial function of $n$. 

We may assume without loss of generality that all predicates in
$\Delta$ occur in $\varphi$ or $\psi(\bar{y})$, since---provided
$\Delta$ contains no direct contradictions---literals involving
foreign predicates can simply be removed. Further, we may assume that,
for every ground atom $\alpha$ over the relevant signature, $\Delta$
contains either $\alpha$ or $\neg \alpha$. For if not,
non-deterministically add either of these literals to $\Delta$; since
all predicates of $\varphi$ and $\psi(\bar{y})$ are by hypothesis of
arity 1 or 2, this process may be carried out in time bounded by a
quadratic function of $n$. Finally, we may assume that, for all
distinct $c, d \in {\rm const}(\Delta) \cup \bar{a}$, $\Delta$
contains the literal $c \not \approx d$, since, if $\Delta$ contains
$c \approx d$, either of these constants can be eliminated.

Since $\psi(\bar{y})$ is a positive conjunctive query, we may take
$\neg \psi(\bar{a})$ to be $\forall \bar{x} \eta$, where $\eta$ is a
negative clause. Let $K = {\rm const}(\Delta) \cup \bar{a}$, and let
$\Xi$ be the set of functions from $\bar{x}$ to $\bar{x} \cup
K$. Thus, $|\Xi| \leq (n+l_1 +l_2)^{l_1}$, where $l_1$ is the arity of
$\bar{x}$ and $l_2$ is the arity of $\bar{y}$.  
Employing the notation of Lemma~\ref{lma:conjunctionOverAssignments},
and recalling Definition~\ref{def:cyclic}, let
\begin{eqnarray*}
\Xi_1 & = & \{ \xi \in \Xi \mid \eta_\xi \mbox{ is v-cyclic} \}\\
\Xi_2 & = & \{ \xi \in \Xi \mid \eta_\xi \mbox{ is v-acyclic} \}.
\end{eqnarray*}
Thus, Formula~\eqref{eq:coProblem} is logically equivalent to
\begin{equation}
\bigwedge \Delta \wedge \varphi 
\wedge
\bigwedge_{\xi \in \Xi_1} \forall \bar{x}_\xi 
       \big(\big(
        \bigwedge_{\substack{x \in \bar{x}_\xi\\
                             c  \in K}} 
                             x \not \approx c
        \wedge
        \bigwedge_{\substack{x, x'  \in \bar{x}_\xi\\
                             x  \neq x'}} 
               x \not \approx x'\big) \rightarrow \eta_\xi \big)
\wedge
\bigwedge_{\xi \in \Xi_2} \forall \bar{x}_\xi 
       \eta_\xi; 
\label{eq:halfWay}
\end{equation}
moreover, this latter formula can be computed in time bounded by a
polynomial function of $|\Xi|$, and hence of $n$. Let us
write~\eqref{eq:halfWay}
as 
\begin{equation}
\bigwedge \Delta \wedge \varphi 
\wedge 
\bigwedge \Upsilon
\wedge
\bigwedge_{\xi \in \Xi_2} \forall \bar{x}_\xi 
       \eta_\xi; 
\label{eq:halfWayBis}
\end{equation}
where $\Upsilon$ is a finite set of v-formulas with respect to
$K$. Let $\eta_\xi^\Delta$ denote $\top$ if any ground literal of
$\eta_\xi$ appears in $\Delta$; otherwise, let $\eta_\xi^\Delta$ be
the result of deleting from $\eta_\xi$ all ground literals whose
negation appears in $\Delta$. (If no literals remain, $\eta^\Delta$ is
taken to be $\bot$.) Thus, \eqref{eq:halfWayBis} is logically
equivalent to
\begin{equation}
\bigwedge \Delta \wedge \varphi 
\wedge 
\bigwedge \Upsilon
\wedge
\bigwedge_{\xi \in \Xi_2} \forall \bar{x}_\xi 
       \eta^\Delta_\xi.
\label{eq:halfWayBisBis}
\end{equation}
Since $\Delta$ contains every ground literal or its negation over the
relevant signature, no ground literal can appear in any of the
$\eta^\Delta_\xi$. Moreover, if any of the $\eta^\Delta_\xi$ is empty,
\eqref{eq:halfWayBisBis} is trivially unsatisfiable; so we may suppose
otherwise. List the formulas $\forall \bar{x}_\xi \eta_\xi^\Delta$ for
$\xi \in \Xi_2$, as $\forall \bar{x}_i \eta_i$ ($1 \leq i \leq s$);
and re-write each $\forall \bar{x}_i \eta_i$ as a disjunction
\begin{equation*}
\forall \bar{x}_{i,1} \eta_{i,1} \vee \cdots \vee \forall \bar{x}_{i,t_i} \eta_{i,t_i}
\end{equation*}
where the $\eta_{i,j}$ are unsplittable. For each $i$ ($1 \leq i \leq
s$), pick a value $j$ ($1 \leq j \leq t_i$) and write $\forall
\bar{x}_{i,j} \eta_{i,j}$ as $\forall \bar{x}'_i \eta'_i$. Thus,
\eqref{eq:halfWayBisBis} is finitely satisfiable if and only if, for some
way of making the above choices, the resulting formula
\begin{equation}
\bigwedge \Delta \wedge \varphi 
\wedge
\bigwedge \Upsilon
\wedge
\bigwedge_{1 \leq i \leq s} \forall \bar{x}'_i \eta'_i 
\label{eq:halfWayAndABit}
\end{equation}
is finitely satisfiable. This (non-deterministic) step may again be
executed in time bounded by a polynomial function of $n$.  Note that
each $\eta'_i$ is v-acyclic, unsplittable and non-ground; hence, by
Remark~\ref{remark:unsplittable}, it contains at most one individual
constant. We may assume for simplicity, and without loss of
generality, that $\eta'_i$ contains exactly one individual
constant---say, $c_i$.

Let $\eta''_i$ be the result of replacing all occurrences of $c_i$ in
$\eta'_i$ by $x$ (where $x$ does not occur in $\eta'_i$), and let
$p_i$ be a new unary predicate {\em depending only on the clause
$\eta''_i$} (and not on $i$): that is, if $\eta''_i = \eta''_j$, then
$p_i = p_j$. Since $\eta'_i$ contains at most one individual constant,
$\eta''_i$ is a clause in the signature of $\psi(\bar{y})$; therefore,
the number of distinct predicates $p_i$ is bounded by some constant,
independent of $\Delta$. Let $\Delta' = \{ p_i(c_i) \mid 1 \leq i \leq
s \}$.  By Lemma~\ref{lma:getRidOfConstants}, then,
\eqref{eq:halfWayAndABit} is satisfiable over the same domains as
\begin{equation}
\bigwedge (\Delta \cup \Delta') \wedge \varphi 
\wedge
\bigwedge \Upsilon
\bigwedge_{1 \leq i \leq s} \forall x\bar{x}'_i (\eta''_i \vee \neg p_i(x)).
\label{eq:halfWayAndABitMore}
\end{equation}

Evidently, \eqref{eq:halfWayAndABitMore} can be computed in time
bounded by a polynomial function of $n$; in particular, $|\Delta'|$ is
also bounded in this way. However, the number of formulas $\forall
x\bar{x}'_i (\eta''_i \vee \neg p_i(x))$ occurring
in~\eqref{eq:halfWayAndABitMore}---assuming duplicates to be
omitted---is bounded by a constant.  By Lemma~\ref{lma:pureTformulas},
there exists, for each such $\forall x\bar{x}'_i (\eta''_i \vee \neg
p_i(x))$, a logically equivalent $\cGC^2$-formula $\forall x
\theta_i(x)$. Let $\theta$ be the conjunction of all these $\forall x
\theta_i(x)$. Then~\eqref{eq:halfWayAndABitMore} is logically
equivalent to
\begin{equation}
\bigwedge (\Delta \cup \Delta') \wedge (\varphi \wedge \theta) \wedge \Upsilon.
\label{eq:almostThere}
\end{equation}
Finally, by Lemma~\ref{lma:useBigLoops}, \eqref{eq:almostThere} is
finitely satisfiable if and only if
\begin{equation}
\bigwedge (\Delta \cup  \Delta') \wedge (\varphi \wedge \theta)
\label{eq:noUpsilons}
\end{equation}
is finitely satisfiable. Since $(\varphi \wedge \theta)$ is a one of a
finite number $H$ of possible $\cGC^2$- (and hence $\cC^2$-) formulas,
where $H$ depends only on the signature of $\psi(\bar{y})$, and not on
$\Delta$, the finite satisfiability of~\eqref{eq:noUpsilons} can be
tested nondeterministically in time bounded by a polynomial function
of $n$, by Theorem~\ref{theo:dataSatC2}.
\end{proof}
That the same complexity bounds are obtained for the query-answering
and finite query-answering problems in
Theorem~\ref{theo:queryAnswering} is, incidentally, not something that
should be taken for granted. For example, Rosati~\cite{logic:rosati06}
presents a relatively simple logic (not a subset of $\cC^2$) for which
query-answering is always decidable, but finite query-answering in
general undecidable.  

\paragraph{Acknowledgments}
This paper was written during a visit to the Faculty of Computer
Science at the Free University of Bozen-Bolzano. The author wishes to
express his appreciation for this opportunity and to acknowledge the
help of Diego Calvanese, David Toman and Alessandro Artale in
discussions on this topic.

\bibliographystyle{plain} 
\bibliography{dataC2}
\end{document}